 \documentclass[twocolumn,tighten]{aastex63}
\usepackage{amsmath}

\shorttitle{Testing the Strong Equivalence Principle with Rotationally Supported Galaxies}
\shortauthors{Chae et al.}

\begin{document}

\title{Testing the Strong Equivalence Principle: Detection of the External Field Effect in Rotationally Supported Galaxies}

\correspondingauthor{Kyu-Hyun Chae}

\author{Kyu-Hyun Chae}
\affil{Department of Physics and Astronomy, Sejong University, 209 Neungdong-ro Gwangjin-gu, Seoul 05006, Republic of Korea}
\email{KHC: chae@sejong.ac.kr, kyuhyunchae@gmail.com}

\author{Federico Lelli}
\affil{School of Physics and Astronomy, Cardiff University, Queens Buildings, The Parade, Cardiff, CF24 3AA, UK}
\email{FL: LelliF@cardiff.ac.uk}

\author{Harry Desmond}
\affil{Astrophysics, University of Oxford, Denys Wilkinson Building, Keble Road, Oxford, OX1 3RH, UK}
\email{HD: harry.desmond@physics.ox.ac.uk}

\author{Stacy S. McGaugh}
\affil{Department of Astronomy, Case Western Reserve University, Cleveland, OH 44106, USA}
\email{SSM: ssm69@case.edu}

\author{Pengfei Li}
\affil{Department of Astronomy, Case Western Reserve University, Cleveland, OH 44106, USA}
\email{PL: pxl283@case.edu}

\author{James M. Schombert}
\affil{Department of Physics, University of Oregon, Eugene, OR 97403, USA}
\email{JMS: jschombe@gmail.com}



\begin{abstract}
  The strong equivalence principle (SEP) distinguishes General Relativity from other viable theories of gravity. The SEP demands that the internal dynamics of a self-gravitating system under free-fall in an external gravitational field should not depend on the external field strength. We test the SEP by investigating the external field effect (EFE) in Milgromian dynamics (MOND), proposed as an alternative to dark matter in interpreting galactic kinematics. We report a detection of this EFE using galaxies from the Spitzer Photometry and Accurate Rotation Curves (SPARC) sample together with estimates of the large-scale external gravitational field from an all-sky galaxy catalog. Our detection is threefold: (1) the EFE is individually detected at $8\sigma$ to $11\sigma$ in ``golden'' galaxies subjected to exceptionally strong external fields, while it is not detected in exceptionally isolated galaxies, (2) the EFE is statistically detected at more than $4\sigma$ from a blind test of 153 SPARC rotating galaxies, giving a mean value of the external field consistent with an independent estimate from the galaxies' environments, and (3) we detect a systematic downward trend in the weak gravity part of the radial acceleration relation at the right acceleration predicted by the EFE of the MOND modified gravity. Tidal effects from neighboring galaxies in the $\Lambda$CDM context are not strong enough to explain these phenomena. They are not predicted by existing $\Lambda$CDM models of galaxy formation and evolution, adding a new small-scale challenge to the $\Lambda$CDM paradigm. Our results point to a breakdown of the SEP, supporting modified gravity theories beyond General Relativity.
\end{abstract}




\section{Introduction} \label{sec:intro}

The hypothesis that General Relativity (GR) and its Newtonian limit hold exactly in the weak gravity regime requires that the Universe is permeated by invisible dark matter (DM). The existence of DM is a key assumption of the standard cosmological model $\Lambda$ Cold Dark Matter ($\Lambda$CDM), which has been successful in explaining many cosmological observations on the largest scales of the cosmos \citep{Peb12, FW12}. The $\Lambda$CDM paradigm, however, is facing several challenges on small scales \citep{BBK17, Krou15}, such as the unexpected phase-space correlation of satellite galaxies (``the satellite plane problem''; see, e.g., \citealt{Krou10,Mul18}) and the unexpected coupling in galaxies between the visible matter (baryons) and the observed dynamics, usually dominated by the DM halo at large radii \citep{MLS, Lel17}.

A drastically different idea is represented by the MOND paradigm \citep{Mil83} that modifies the standard laws of dynamics at low accelerations (weak gravitational fields) rather than assuming non-baryonic DM. Several a-priori predictions of MOND have been confirmed by later observations as reviewed by \cite{SM02}, \cite{FM12}, and \cite{McG20}. The construction of a MOND cosmology remains a tall order \citep{McG15}, but the recent relativistic MOND theory of \cite{Skor20} appears promising, being able to reproduce the power spectrum of the Cosmic Microwave Background as good as $\Lambda$CDM.

The relativistic theory of \cite{Skor20} reduces to the non-relativistic modified-gravity theory of \cite{BM84}, violating the Strong Equivalence Principle (SEP) of GR: the internal dynamics of a self-gravitating body may be affected by external gravitational fields, beyond usual tidal forces. More specifically, these theories violate Local Positional Invariance (LPI) for gravitational experiments, which differentiates the SEP from the less stringent (but well tested) Einstein Equivalence Principle (EEP), containing the Weak Equivalence Principle, Lorentz Invariance, and the LPI for non-gravitational experiments only \citep{Will14}.

The radial acceleration relation (RAR) is of particular importance in the DM vs MOND debate \citep{MLS,Lel17}. This empirical relationship links the observed centripetal acceleration $g_{\rm obs}(R)=V^2_{\rm rot}(R)/R$ in galaxies to the expected Newtonian acceleration $g_{\rm bar}(R) =V^2_{\rm bar}(R)/R$ from the observed baryonic matter distribution:
\begin{equation}
  g_{\rm obs} = \nu_0 \left(\dfrac{g_{\rm bar}}{g_\dagger}\right) g_{\rm bar}
  \label{eq:rar}
\end{equation}
where $\nu_0(z)$ is an empirical fitting function and $g_\dagger$ is an acceleration scale. In $\Lambda$CDM the RAR must arise from the haphazard process of galaxy formation \citep{DCL16, Des17, Nav17, KW17} and $g_\dagger$ is an emergent scale that may \citep{Lud17} or may not\citep{Ten18} appear in cosmological simulations. In MOND $g_\dagger$ is a new universal constant of Nature indicated as $a_0$ \citep{Mil83}, while the function $\nu_0(g_{\rm bar}/a_0)$ interpolates between the classic Newtonian regime $g_{\rm obs} = g_{\rm bar}$ at high accelerations and the Milgromian regime $g_{\rm obs} = \sqrt{g_{\rm bar}a_0}$ at low accelerations. 

While the extrapolation of Eq.~(\ref{eq:rar}) to large radii implies asymptotically flat rotation curves for isolated galaxies, MOND modified-gravity \citep{BM84} predicts that galaxies in strong external fields should display a weak but distinctive decline in their outer rotation curves. This peculiar feature, linking the internal dynamics on scales smaller than 100 kpc with the cosmological environment on scales of a few Mpc, can be used to distinguish between modified gravity in MOND and standard gravity with DM. Signatures of this external field effect (EFE) have been searched for in rotationally-supported galaxies \citep{Hag16,WK15,Lel15} without conclusive and unambiguous evidence.

The EFE has also been investigated in pressure-supported stellar systems. Dwarf satellites of the Andromeda galaxy revealed some EFE signatures as predicted and tested by \cite{MM13a,MM13b}, but the possibility of tidal interactions and out-of-equilibrium dynamics complicates the interpretation (e.g.\ \citealt{MW10,Lel17}). Several authors \citep{FMM18,Krou18,Mul19,Hag19} proposed MOND models incorporating the EFE to explain unexpectedly low stellar velocity dispersions of a few ultra-diffuse galaxies. Globular clusters (GCs) of the Milky Way are dynamical systems subjected to external fields. MONDian kinematics for the GCs were predicted \citep{Baum05,Hag09,Hag11}, but analyses of the observed data did not result in unambiguous signatures of the MOND EFE \citep{Jor09,Fra12}.

Wide binary stars have also been used to test MOND and the EFE, with conflicting results \citep{Her12,PS19,Her19}. In particular, wide binary stars from GAIA DR2 have been used to argue both for \citep{PS19} and against \citep{Her19} the presence of the EFE, and further studies are required to provide conclusive evidence.

Here we report a robust EFE detection in rotationally supported galaxies using two complementary approaches: (1) focusing on individual galaxies where the external gravitational field is exceptionally large, (2) studying weak systematic deviations from the RAR driven by the mean gravitational field of the Local Universe. Throughout we take $g_\dagger=1.2\times 10^{-10}$ m~s$^{-2}$ \citep{MLS,Lel17} and use the notation $x\equiv \log_{\rm 10} (g_{\rm bar}/{\rm m~s^{-2}})$ and $y\equiv \log_{\rm 10} (g_{\rm obs}/{\rm m~s^{-2}})$.

\section{Data and Methodology} \label{sec:method}

\subsection{The SPARC database} \label{sec:sparc}

The SPARC database \citep{Lel16} contains 175 rotationally-supported galaxies in the nearby Universe\footnote{\mbox{http://astroweb.cwru.edu/SPARC/}}. These galaxies have stellar masses ranging from $M_\star \simeq 10^{11} M_\odot$ to $M_\star \simeq 10^{7} M_\odot$ and cover all Hubble types of late-type, star-forming galaxies, including low-surface brightness disk galaxies. The database provides the observed rotation velocities ($V_{\rm obs}$) from spatially resolved HI observations and the Newtonian circular velocities from the observed distribution of stars and gas. The latter include the stellar disk contribution ($V_{\rm disk}$) and (if present) the bulge contribution ($V_{\rm bul}$) for a baseline mass-to-light ratios of unity, as well as the gas contribution ($V_{\rm gas}$) for a total-to-hydrogen mass ratio of $1.33$. For convenience, in this paper we redefine $V_{\rm gas}$ for a total-to-hydrogen mass ratio of unity. The reported velocity $V_{\rm obs}$ of a galaxy is tied to the reported inclination $i_{\rm obs}$. If the inclination is changed to $i$, the rotation velocity becomes
  \begin{equation}
    V_{\rm rot}=V_{\rm obs} \frac{\sin(i_{\rm obs})}{\sin(i)}.
    \label{eq:Vrot}
    \end{equation}

The circular velocity due to the baryonic mass distribution depends on the galaxy distance $D$ and is given by
  \begin{equation}
    V_{\rm bar}=\sqrt{\hat{D}(\Upsilon_{\rm disk}V_{\rm disk}^2+\Upsilon_{\rm bul}V_{\rm bul}^2+\Upsilon_{\rm gas}V_{\rm gas}|V_{\rm gas}|)},
    \label{eq:Vbar}
    \end{equation}
where $\hat{D} \equiv D/D_{\rm obs}$ with $D_{\rm obs}$ being the fiducial distance. In Eq.~(\ref{eq:Vbar}), $\Upsilon_{\rm disk}$ and $\Upsilon_{\rm bul}$ are the mass-to-light ratios of the disk and the bulge in units of the solar value $M_{\odot}/L_{\odot}$ at $3.6\mu m$, while $\Upsilon_{\rm gas}$ is the ratio of the total gas mass to the HI mass. When the SPARC database was published, this ratio was assumed to be $1.33$ to account for the cosmic abundance of Helium from big bang nucleosynthesis. Here we consider the small amounts of Helium and metals formed via stellar nucleosynthesis during galaxy evolution \citep{MLS20}, so that $\Upsilon_{\rm gas} = X^{-1}$ where $X$ is a function of stellar mass ($M_\star$):
  \begin{equation}
    X=0.75-38.2 \left( \frac{M_{\star}}{M_0} \right)^\alpha,
     \label{eq:X1}
  \end{equation}
with $M_0=1.5\times 10^{24}{\rm M}_\odot$ and $\alpha=0.22$. We do however allow the possibility of varying $\Upsilon_{\rm gas}$ from $X^{-1}$ to consider the uncertainties in the HI flux, gas disk geometry, and the gas mass to HI mass ratio. In some cases $V_{\rm gas}$ is negative at small radii, representing the fact that the Newtonian gravitational field is not oriented towards the center when a large fraction of the gas disk lies in the outer regions. To account for the cases of negative $V_{\rm gas}$ we write $V_{\rm gas}|V_{\rm gas}|$ rather than $V_{\rm gas}^2$ in the last term of Equation~(\ref{eq:Vbar}), although this detail has negligible effects on our study.

\subsection{The external field effect}  \label{sec:efe}

Empirically, the observed centripetal acceleration ($g_{\rm obs} = V_{\rm obs}^2/R$) is related to the Newtonian baryonic acceleration ($g_{\rm bar} = V^{2}_{\rm bar}/R$) via the RAR $\nu_0(g_{\rm bar}/g_\dagger)$ of Eq.~(\ref{eq:rar}) with a free parameter $g_\dagger$ \citep{MLS, Lel17}. In a MOND framework, $g_\dagger=a_0$ is a fundamental constant of Nature \citep{Mil83} and Eq.~(\ref{eq:rar}) can be obtained by modifying either inertia (Newton's second law of dynamics) or gravity (the Poisson's equation) at the non-relativistic level \citep{FM12}. In MOND modified-inertia theories Eq.~(\ref{eq:rar}) holds exactly for any circular orbit \citep{Mil94}, while in MOND modified gravity theories holds only for highly symmetric mass distributions (such as spheres) and represents a first-order approximation for actual disk galaxies \citep{BM95}. In all these scenarios, however, Eq.~(\ref{eq:rar}) is strictly valid only for isolated systems, when the external field effect (EFE) is negligible.

\begin{figure}
  \centering
  \includegraphics[width=1.\linewidth]{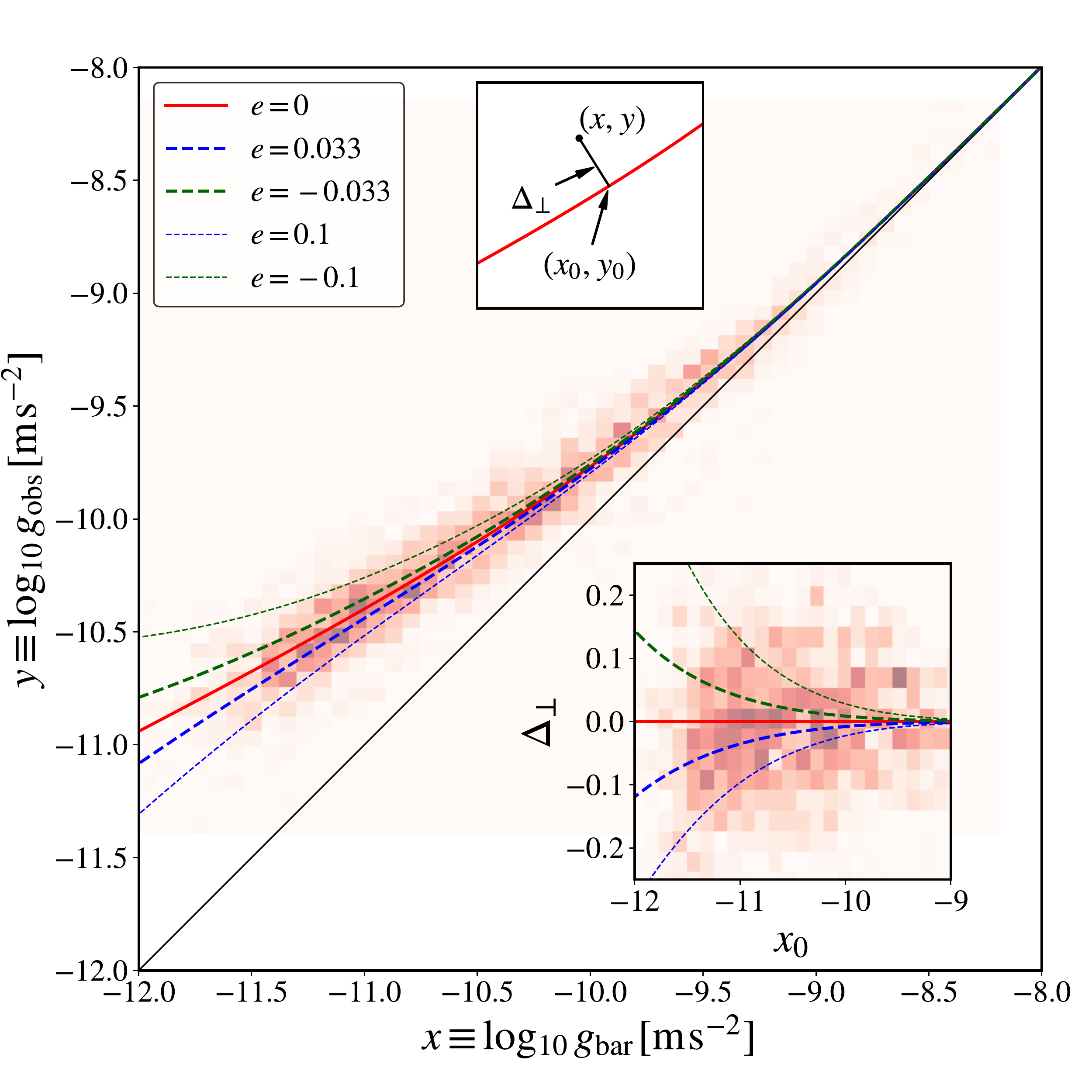}
    \vspace{-0.5truecm}  
    \caption{\small 
    The external field effect in the weak-field limit of the radial acceleration relation.
 Eq.~(\ref{eq:rare}) is overlaid on the RAR for various values of $e$ in Eq.~(\ref{eq:rare}). Values of $e>0$ correspond to the MOND EFE, while $e<0$ is unphysical from the MOND point of view. Values of $e\approx 0.033$ corresponds to the average prediction for 153 SPARC galaxies based on their gravitational environments \citep{Des18}. The heat map shows the original SPARC mass models \citep{Lel16} with fixed stellar mass-to-light ratios for the same galaxies.}
   \label{rar_theory}
\end{figure}

To build a general fitting function that approximates the EFE, we start from the nonlinear MOND modified Poisson's equation \citep{BM84} in the one-dimensional case. If we assume a uniform external gravitational field $g_{\rm ext} $\citep{FM12} and the so-called Simple interpolating function (IF) \citep{FB05}, we have
\begin{equation}
g_{\rm MOND}(R) = \nu_e\left( \frac{g_{\rm bar}}{g_\dagger} \right) g_{\rm bar}(R)
  \label{eq:rare}
\end{equation}
with
\begin{equation}
  \nu_e(z) = \frac{1}{2}- \frac{A_e }{z} + \sqrt{\left(\frac{1}{2}- \frac{A_e}{z}\right)^2 +  \frac{B_e}{z}},
  \label{eq:nue}
\end{equation}  
where $z \equiv g_{\rm bar}/g_\dagger$, $A_e \equiv e(1+e/2)/(1+e)$, $B_e\equiv(1+e)$, and $e \equiv g_{\rm ext}/g_\dagger$. For $e=0$, $\nu_e(z)$ is reduced to the Simple IF $\nu_0(z) = 1/2 + \sqrt{1/4+1/z}$. Equation~(\ref{eq:nue}) is based on the footnote to Eq.\ (59) of \cite{FM12}, but we corrected a small typo and rearranged it.  Note here that the Simple IF allows the convenient analytic form of Equation~(\ref{eq:nue}) with $e>0$ while it is only subtly different \citep{Chae19} in the (EFE-irrelevant) high acceleration limit from the function used by \cite{MLS} and \cite{Lel17} to fit the SPARC galaxies. Our results on the EFE detection are not affected by the choice of the Simple IF. Then, the expected circular velocity is given by
\begin{equation}
V_{\rm MOND}(R)=\sqrt{\nu_e\left( \frac{g_{\rm bar}}{g_\dagger} \right)} V_{\rm bar}(R).
\label{eq:Vtot}
\end{equation}

Although $\nu_e(z)$ (Eq.~\ref{eq:nue}) is based on idealized assumptions, it captures the basic feature of the EFE: a systematic downward deviation from $\nu_0(z)$ (Eq.~\ref{eq:rar}) when $e>0$ as $z \rightarrow 0$. Eq.~(\ref{eq:nue}) also allows for upward deviations when $e<0$, which seem unphysical but may be preferred by the data at the empirical level. These features are illustrated in Fig.~\ref{rar_theory}. MOND with the EFE predicts that the RAR must be a family of functions rather than a universal function. This also means that if galaxies in different environments are tried to be fitted with a single functional form of Eq.~(\ref{eq:rar}), then there will arise some small intrinsic scatter of $g_\dagger$ due to the EFE.  Most importantly, regardless of its MOND origin, Eq.~(\ref{eq:nue}) may be considered a mere fitting function that improves over Eq.~(\ref{eq:rar}) by adding the free parameter $e$, which has no a-priori knowledge of the external gravitational field in which galaxies reside.

\subsection{MCMC simulations}  \label{sec:mcmc}

In our Bayesian analysis the posterior probability of parameters $\vec{\beta}=\{\beta_k\}$ is defined by
   \begin{equation}
     p(\vec{\beta}) \propto \exp\left( -\frac{\chi^2}{2} \right) \prod_k {\rm Pr}(\beta_k),
     \label{eq:pdf}
   \end{equation}
where ${\rm Pr}(\beta_k)$ is the prior probability of parameter $\beta_k$ and $\chi^2$ is given by
   \begin{equation}
     \chi^2 = \sum_{j=1}^{N} \left( \frac{V_{\rm rot}(R_j) - V_{\rm MOND}(\vec{\beta};R_j)}{\sigma_{V_{\rm rot}(R_j)}} \right)^2,
     \label{eq:chi2}
   \end{equation}
with $\sigma_{V_{\rm rot}(R_j)}= \sigma_{V_{\rm obs}(R_j)} \sin(i_{\rm obs})/\sin(i)$ where $\sigma_{V_{\rm obs}(R_j)}$ is the reported error of $V_{\rm obs}(R_j)$ for the reported inclination $i_{\rm obs}$. As in earlier studies of the RAR using SPARC galaxies \citep{MLS, Lel17}, we use only 153 galaxies with $i_{\rm obs} \ge 30^\circ$ and $Q \le 2$ (a quality cut on the rotation curve).

The parameters $\vec{\beta}$ in Eq.~(\ref{eq:pdf}) are given by $\vec{\beta}=\{\Upsilon_{\rm disk}, \Upsilon_{\rm bul}, \Upsilon_{\rm gas}, \hat{D}, i, e\}$ for the case of using Eq.~(\ref{eq:rare}) with a fixed $g_\dagger=1.2\times 10^{-10}$~m~s$^{-2}$. The priors on these parameters are summarized in Table~1. The mean values and standard deviations of $\Upsilon_{\rm disk}$ and $\Upsilon_{\rm bul}$ are motivated by state-of-the-art stellar population synthesis models for star-forming galaxies \citep{Schom19}. The mean value of $\Upsilon_{\rm gas}$ is given by Eq.~(\ref{eq:X1}), while the standard deviation is motivated by the typical error on the HI flux calibration, but it could also represent variations in the assumed gas disk thickness and/or the mean gas-to-HI mass ratio. The mean values and standard deviations of $\hat{D}$ and $i$ consider the baseline SPARC values and their fiducial errors. For $e$ we adopt an uninformative uniform prior covering a reasonably broad range.

\begin{table}
\caption{Summary of prior constraints on the model parameters}
\begin{center}
  \begin{tabular}{lcc} \hline
 parameter  &  Distribution  & ($\mu$, $\sigma$) or range \\
 \hline
 $\Upsilon_{\rm disk}$     & Lognormal  & ($\log_{10}(0.5)$, $0.1$)     \\
 $\Upsilon_{\rm bul}$      & Lognormal  & ($\log_{10}(0.7)$, $0.1$)  \\
 $\Upsilon_{\rm gas}$      & Lognormal  &  ($\log_{10}(X^{-1})$, $0.04$)     \\
 $\hat{D}$                 & Lognormal  & ($0$, $\log_{10}(1+\sigma_{D_{\rm obs}}/D_{\rm obs})$)      \\
 $i$                       & Gaussian   & ($i_{\rm obs}$, $\sigma_{i_{\rm obs}}$)    \\
 $e$                       &  Uniform   &  $[-0.5,0.5]$\\
 \hline
\end{tabular}
\end{center}
\end{table}

The posterior probability density functions (PDFs) of the model parameters are derived from MCMC simulations through the public code {\tt emcee} \citep{emcee}. These simulations represent an extension to the previous SPARC analysis \citep{Li18} including the EFE parameter $e$. We choose $N_{\rm walkers}=10000$ and $N_{\rm iteration}=6000$. We discard models up to $N_{\rm iteration}=500$ and thin the rest by a factor of 50 as the auto-correlation lengths for the parameters are $<100$. The posterior PDFs of $x = \log_{10} g_{\rm bar}(R)$ and $y = \log_{10} g_{\rm tot}(R)$ follow from the posterior PDFs of the parameters $i$, $\log_{10} \hat{D}$, $\log_{10} \Upsilon_{\rm disk}$, $\log_{10} \Upsilon_{\rm bul}$, and $\log_{10} \Upsilon_{\rm gas}$.

\subsection{The environmental gravitational field}  \label{sec:genv}

We estimate the environmental gravitational field $g_{\rm env}$ due to the large-scale distribution of matter at the positions of the SPARC galaxies. We perform this calculation within the standard $\Lambda$CDM context \citep{Des18}. A similar calculation is not feasible in a MOND context due to the strong non-linearities in the theory and the lack of a proper MOND cosmology. The $\Lambda$CDM calculation, however, is a good first-order approximation for MOND and other modified gravity theories \citep{Des18}, up to some systematic uncertainty due to the unknown relation between $g_{\rm env}$ in these theories. We use $g_{\rm env}$ primarily for the purpose of picking out extreme cases with exceptionally high or low $g_{\rm env}$ (which should remain true in a relative sense in any cosmological scenario) and to check that the maximum-likelihood values of $e$ from fitting Eq.~(\ref{eq:rare}) are sensible in an order-of-magnitude fashion.

Our calculation of $g_{\rm env}$ starts with the total dynamical masses of the galaxies in the all-sky 2M++ survey \citep{LH11} using abundance matching. We then use N-body simulations in $\Lambda$CDM to populate the surrounding regions with halos hosting galaxies too faint to be recorded in 2M++, using statistical correlations between halo abundances and properties of the galaxy field. Finally, we add mass in long-wavelength modes of the density field according to the inferences of the BORG algorithm \citep{LJ16} applied to the 2M++ catalog. We use the final density field to calculate a posterior distribution for $g_{\rm env}$ at the position of each SPARC galaxy, fully propagating uncertainties in the input quantities. We define $e_{\rm env} \equiv g_{\rm env}/g_\dagger$, and find values in the range $0.01 \la e_{\rm env} \la 0.1$ with a mean of $0.033$ among the SPARC galaxies: typical values are in the range $0.02-0.05$.

\section{Results} \label{sec:res}

\subsection{Individual galaxies} \label{sec:res_ind}

The estimated values of the environmental gravitational field strength $e_{\rm env}$ (\S~\ref{sec:genv}) span almost one order of magnitude, thus there is sufficient dynamic range to check whether the rotation-curve shapes depend or not on the large-scale environment. Among the SPARC galaxies whose rotation curves (RCs) reach $g_{\rm obs} < g_\dagger$, NGC5033 and NGC5055 live in exceptionally dense environments with $e_{\rm env} \approx 0.1$, while NGC1090 and NGC6674 are exceptionally isolated with $e_{\rm env}\approx 0.01$. The former two represent ``golden galaxies'' for the EFE to be detected, while the latter two are control targets for the null detection.

\begin{figure*}
  \centering
  \includegraphics[width=0.75\linewidth]{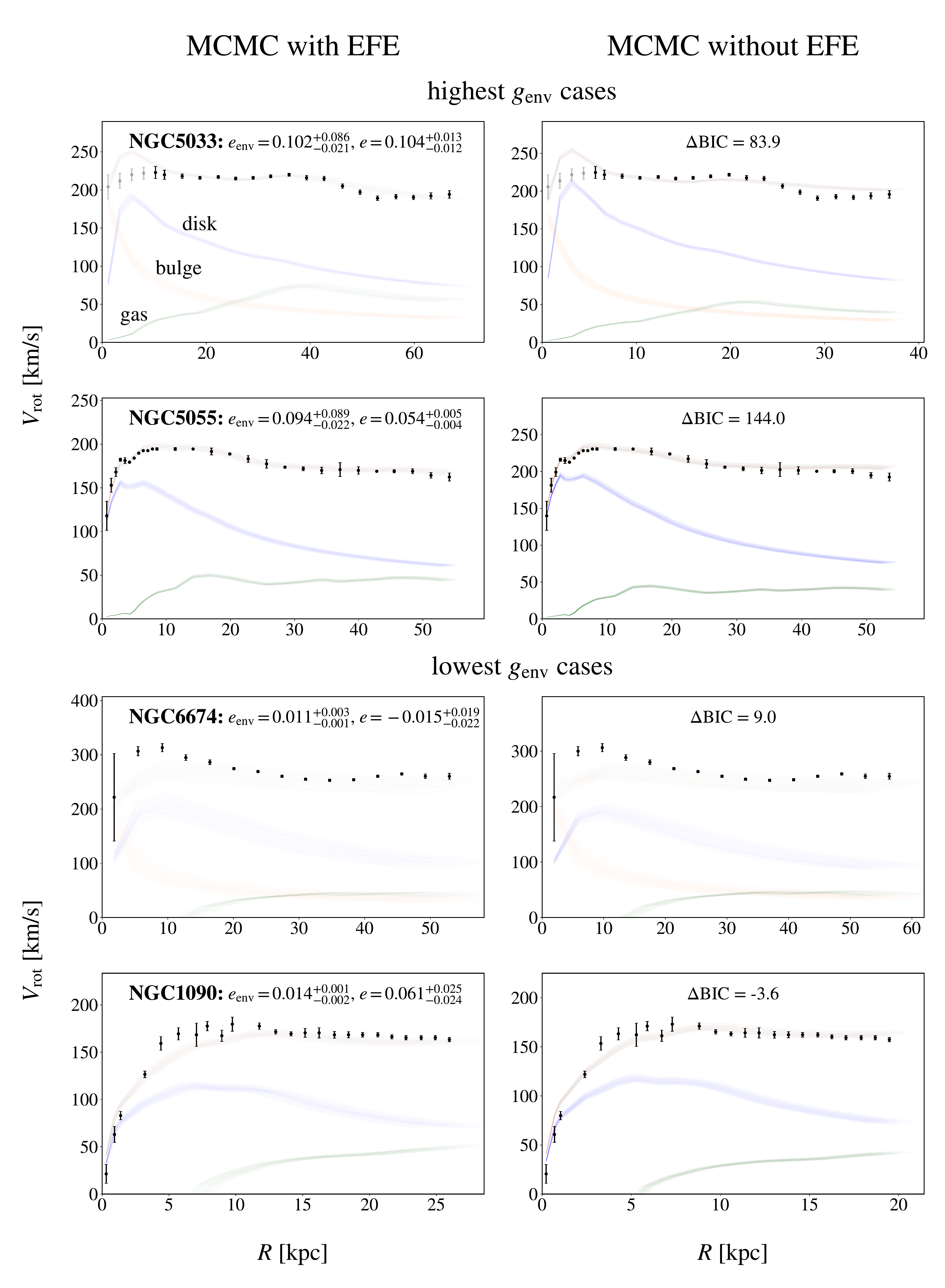}
    \vspace{-0.5truecm}  
    \caption{\small Detection of the EFE in individual galaxies.  The observed rotation curves (points with errorbars) are fitted using Eq.~(\ref{eq:rar}) with no EFE (right panels) and a generalized equation considering the EFE (Equation~\ref{eq:rare}) (left panels). The colored bands show the $1\sigma$ confidence limits for the rotation curve (red) and the separate contributions of gas disk (green), stellar disk (blue), and stellar bulge (orange) if present. For the ``golden galaxies'' subjected to the strongest environmental gravitational fields, the fit is improved dramatically with $e>0$, resulting in $11\sigma$ and $8\sigma$ individual detections of the EFE. For the galaxies subjected to the weakest fields, the EFE is not detected as expected. In all cases, the fitted values of $e$ are fully consistent with the independent values of $e_{\rm env}$ from the large-scale galaxy environment within $\sim 2\sigma$. $\Delta$BIC indicates evidence by the Bayesian Information Criterion.}
   \label{RCgolden}
\end{figure*}

We fit the RCs using the EFE-incorporated RAR fitting function (Eq~\ref{eq:rare}) with a free external field $g_{\rm ext}$ parameterized by $e = g_{\rm ext}/g_\dagger$ (\S~\ref{sec:efe}). The case of $e=0$ implies flat RCs and reduces exactly to the original RAR (Eq.~\ref{eq:rar}). Fig.~\ref{RCgolden} shows the MCMC results for the RCs of four galaxies in the two extreme environments. The `corner' plots showing the posterior PDFs of the parameters for these galaxies can be found in the appendix.

For NGC5055, the detailed shape of the RC is very well fitted with a positive $e$ but poorly fitted with $e=0$. We find $e=0.054\pm 0.005$: this is an $11\sigma$ detection. Remarkably, this value is consistent within 2$\sigma$ with $e_{\rm env}=0.094_{-0.022}^{+0.089}$ that is independently determined from the large-scale environment. The Bayesian information criterion (BIC $\equiv -2\ln L_{\rm max} + k\ln N $ where $L_{\rm max}$ is the maximum likelihood, $k$ is the number of free parameters, and $N$ is the number of the fitted rotation velocities) for $e=0$ relative to the free $e$ case is $\Delta{\rm BIC}=144$, indicating very strong evidence for $e>0$ based on the conventional criterion of $\Delta{\rm BIC}>10$ for strong evidence.

For NGC5033, the overall fit is also improved by freeing up $e$ since $\Delta{\rm BIC}=83.9$. We find $e=0.104_{-0.012}^{+0.013}$. This is an $8\sigma$ detection, in excellent agreement with $e_{\rm env}=0.102_{-0.021}^{+0.086}$ from the large-scale environment. The observed properties of this galaxy, however, are not as robust as those of NGC5055. The rotation velocities at $R<60$~arcseconds (about 5 kpc) are probably underestimated due to beam-smearing effects in the HI data, although our results on $e$ are not affected by these data points. Moreover, while the distance of NGC5055 is robust because it is based on the tip magnitude of the red giant branch ($D=9.90\pm0.30$), that of NGC5033 is very uncertain because it is estimated using Hubble flow models ($D = 15.70  \pm 4.70$). Interestingly, our MCMC result for NGC5033 predicts a relatively large distance ($D=23.5^{+2.0}_{-1.8}$~Mpc) with $e>0$ but a low one ($D=13.0^{+0.7}_{-0.6}$~Mpc) with $e=0$. Hence, future observations can provide a key independent test.

In striking contrast to the highest $e_{\rm env}$ sample, the galaxies in the lowest $e_{\rm env}$ sample show no strong evidence for $e > 0$ based on $\Delta$BIC (or any other widely-used statistic). These two galaxies are similar to the golden galaxies in morphology, mass, and size. The only noticeable difference is that they are unusually isolated. The fitted $e$ values are consistent with the independent $e_{\rm env}$ values within about $2\sigma$.

\subsection{Statistical approach}  \label{sec:res_stat}

Since the EFE has subtle effects on rotation-curve shapes, positive values of $e$ are detected with high statistical significance only in individual galaxies where $e_{\rm env}$ is exceptionally large (like NGC5055 and NGC5033). The EFE, however, should also imprint a statistical signature in the low-acceleration portion of the RAR (see Fig.~\ref{rar_theory}).

\subsubsection{The systematic trend in the low-acceleration portion of the RAR} \label{sec:rardev}

\begin{figure*}
  \centering
  \includegraphics[width=0.8\linewidth]{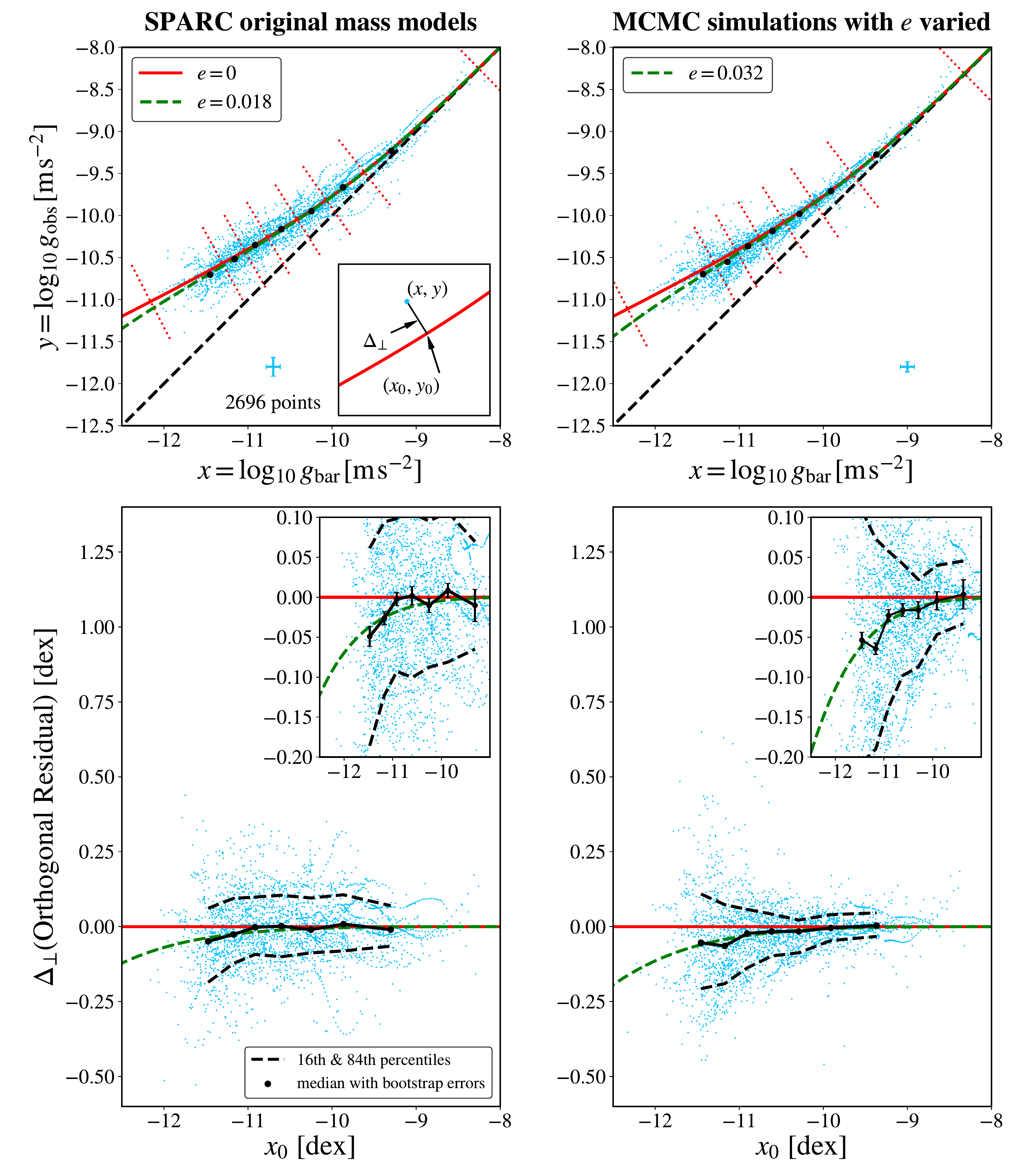}
    \caption{\small 
      EFE detection in the low-acceleration portion of the RAR.
      In the top panels, the Newtonian acceleration from the baryons $g_{\rm bar}$ is plotted against the observed acceleration $g_{\rm obs}$ for a sample of 153 SPARC galaxies. The typical error bars are indicated in the bottom. The data are fitted using Eq.~\ref{eq:rar} (red solid curve with $g_\dagger = 1.2 \times 10^{10}$ m s$^{-2}$) corresponding to $e=0$, and using the additional free parameter $e$ accounting for the EFE (green dashed curve). The black dots show the median values within the bins orthogonal to Eq.~(\ref{eq:rar}) (red dotted lines). The inset illustrates how orthogonal residuals are calculated. The bottom panels show the orthogonal residuals versus $x_0$: the deviation at $x_0 < -11$ represents a statistical detection of the EFE. The inset zooms into this interesting region. The left column shows the original SPARC mass models with fixed stellar mass-to-light ratios, while the right column shows the MCMC results with varied stellar mass-to-light ratios and considering the EFE. In both cases, the fitted $e$ value is remarkably similar to $\langle e_{\rm env}\rangle \approx 0.033$ from the large-scale mass distribution in the nearby Universe.}
   \label{rar}
\end{figure*}

We use 153 galaxies from the SPARC database (\S~\ref{sec:sparc}). Fig.~\ref{rar} (top panels) shows the RAR for 2696 points having accuracy in $V_{\rm rot}$ better than 10$\%$. In the top left panel we first show the original SPARC mass models \citep{Lel16} with fixed mass-to-light ratios at $3.6\mu$m of $\Upsilon_{\rm disk}=0.5 M_\odot/L_\odot$ for the disk and $\Upsilon_{\rm bulge}=0.7 M_\odot/L_\odot$ for the bulge \citep{Lel17}. The MCMC mass models obtained here with varied mass-to-light ratios (\S~\ref{sec:mcmc}) are shown in the top right panel. We divide the data points into bins perpendicular to the best-fit curve assuming Eq.~(\ref{eq:rar}). Each data point $(x,y)$ is projected onto the point $(x_{0},y_{0})$, so the orthogonal residual $\Delta_{\bot}$ encodes any possible systematic deviation from the $e=0$ case.

The data show a small systematic deviation from Eq.~(\ref{eq:rar}) for $x_{0}\lesssim -11$. This trend is present, though weakly, in the original SPARC mass models with fixed mass-to-light ratios for the disks and bulges. The MCMC models in the middle column show a stronger effect. The systematic deviation is weak in absolute terms ($0.05$-$0.08$~dex for the lowest $x_{0}$ bin) but at least 4 times larger than the bootstrap error of the median in the bin. This demonstrates that Eq.~(\ref{eq:rar}) does not fully capture the trends in the RAR. Introducing $e$ as an additional free parameter, we obtain a better fit and find $e\approx 0.02 - 0.04$ in close agreement with the independent estimate of $\langle e_{\rm env}\rangle \simeq 0.033$ from an all-sky galaxy catalog (\S~\ref{sec:genv}).

\subsubsection{The statistical detection of the EFE}  \label{sec:statefe}

The systematic trend in the RAR also implies that the fitted individual values of $e$ of Eq.~(\ref{eq:rare}) will be systematically displaced from the non-EFE case $e=0$. The posterior PDFs of $e$ are quite broad with a typical standard deviation of $\sim 0.04$ (see the appendix for examples). Nevertheless, the statistical distribution of the fitted values will have a signature. Because $e$ was allowed to take any value (positive or negative), this distribution provides a blind test of MOND EFEs (\S~\ref{sec:efe}).

\begin{figure*}
  \centering
  \includegraphics[width=0.9\linewidth]{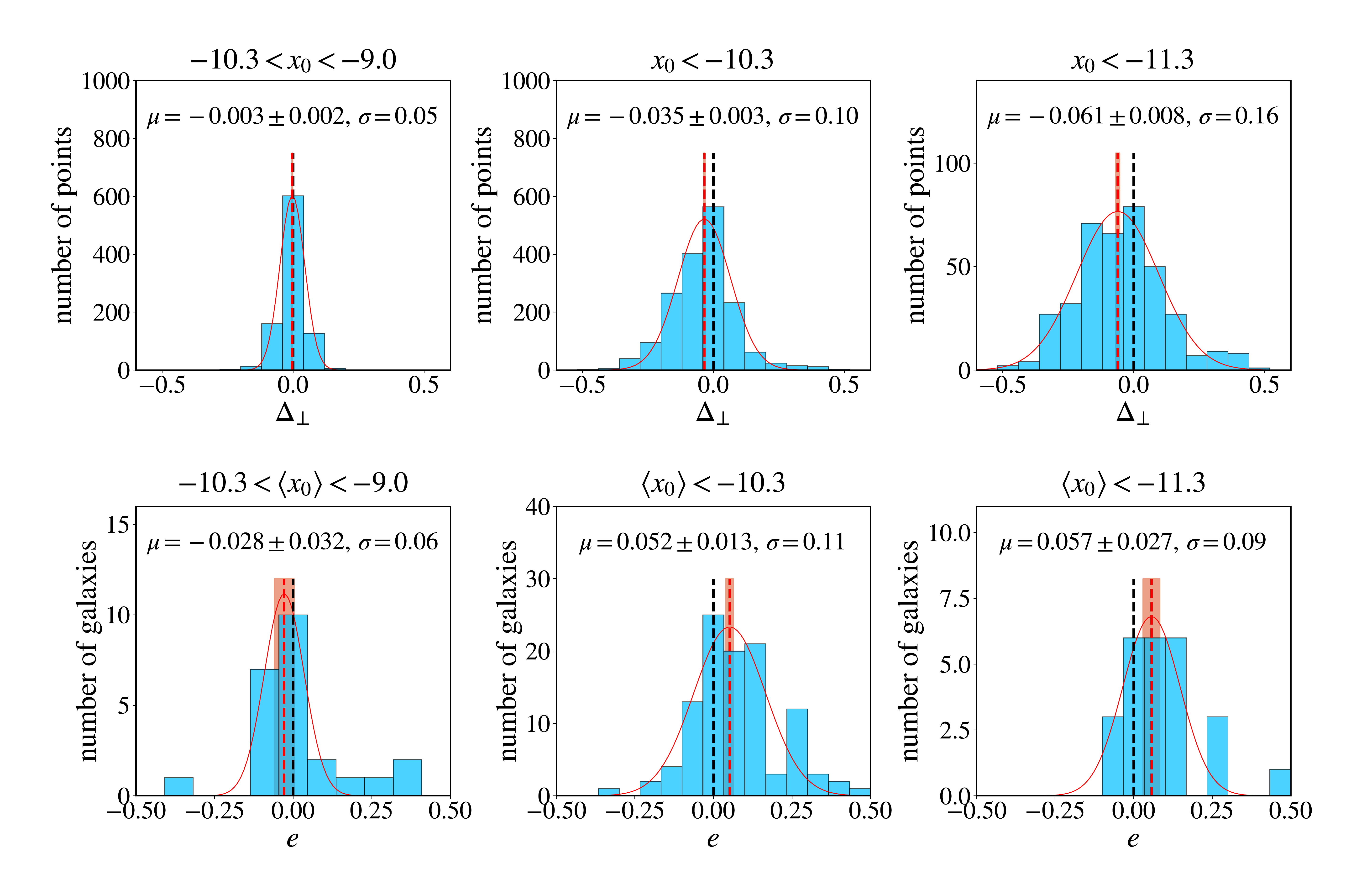}
    \vspace{-0.3truecm}  
    \caption{\small Distributions of $\Delta_{\bot}$ and $e$ from fitting Eq.~(\ref{eq:rare}) to the SPARC galaxies.
      The top panels show the distributions of orthogonal residuals $\Delta_{\bot}$ for three acceleration bins from the MCMC results shown in the middle column of Fig.~\ref{rar}. The mean of the distribution is displaced from zero for lower acceleration bins, indicating declining RCs. The bottom panels show the distributions of the $e$ values fitted to the individual galaxies binned by the median values of $x_{0}$ within the galaxies. As expected, for the galaxies in the high acceleration bin ($-10.3<\langle x_{0}\rangle<-9.0$), the data do not have any sensitivity to $e$ and so the distribution has a mean of $\sim 0$. For lower acceleration bins the distributions are shifted to positive $e$ with high statistical significance, indicating a preference for the EFE. The broad distributions are due to the broad individual posteriors on $e$.   
   }
   \label{gehist}
\end{figure*}

Fig.~\ref{gehist} shows the distribution of the orthogonal residual $\Delta_{\bot}$ and the fitted median value of $e$ from the MCMC simulations with Eq.~(\ref{eq:rare}). From Fig.~\ref{rar_theory} it is expected that data points at high enough accelerations do not have any sensitivity to $e$. Indeed, for data points with $-10.3<x_{0}<-9$ the distribution of $\Delta_{\bot}$ gives a null result. Similarly, for galaxies with $-10.3<\langle x_{0} \rangle <-9$, the distribution of $e_{\rm MLE}$ gives a null result.

Data points at low enough accelerations will have sensitivity to $e$ and  distributions with non-zero mean value are expected from Fig.~\ref{rar}. For data points with $x_{0}<-11.3$ the distribution of $\Delta_{\bot}$ has a mean of $-0.061\pm 0.008$ (a bootstrap error) which is statistically significant at more than $7\sigma$. For a much larger number of data points with $x_{0}<-10.3$, $\Delta_{\bot}$ has a smaller deviation of $-0.035\pm 0.003$, but the statistical significance of the deviation is more than $11\sigma$.

\begin{figure}
  \centering
  \includegraphics[width=1.05\linewidth, trim=25mm 0mm 0mm 20mm, clip]{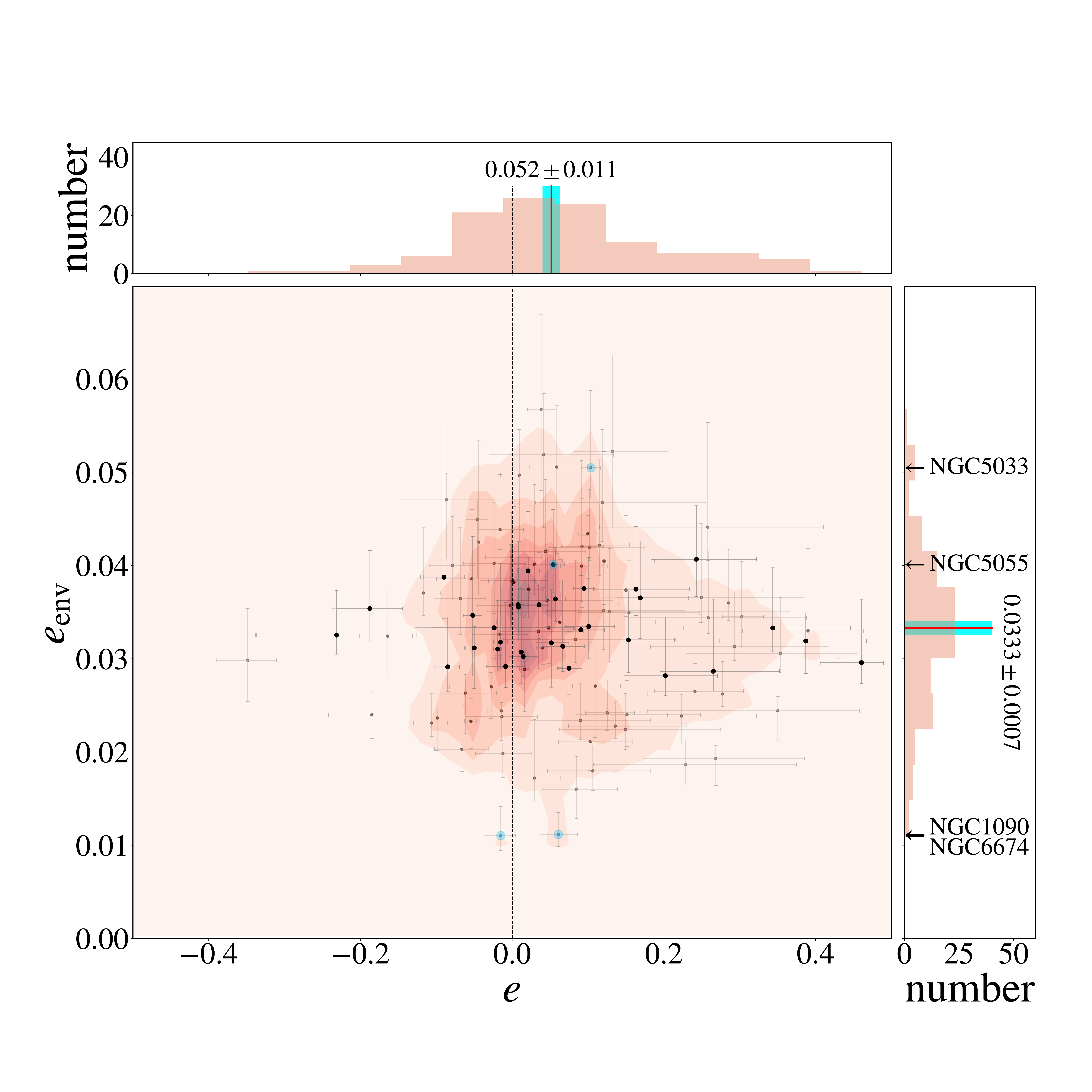}
    \vspace{-1.truecm}  
    \caption{\small Statistical detection of the EFE. 
  The median values of $e$ from rotation-curve fits are compared with $e_{\rm env}$ values from the large-scale galaxy environment \citep{Des18} for 113 galaxies with $\langle x_0 \rangle < -10.3$. The heat map considers the posterior probabilities of individual $e$ measurements. The distributions of $e$ and $e_{\rm env}$ are shown by the top and right histograms. The median value of $e$ is clearly offset from zero, indicating a $5\sigma$ statistical detection of the EFE. The median value of $\langle e \rangle = 0.052\pm 0.011$ is statistically consistent with $\langle e_{\rm env}\rangle = 0.034 \pm 0.001$ (see also Appendix~B). The individual galaxies considered in Fig.~\ref{RCgolden} are indicated: for the golden galaxies at high $e_{\rm env}$ values, $e$ is significantly different from zero at 8$\sigma$ (NGC5033) and 11$\sigma$ (NGC5055). Big dots indicate galaxies with accurate distances.}
  \label{e_values}
\end{figure}

Fig.~\ref{e_values} shows individual $e$ values and their uncertainties for a subset of 113 galaxies with median $\langle x_{0}\rangle < -10.3$. Due to the large uncertainties on $e$, some galaxies can occasionally return negative values. However, the median value of $e$ is $0.052\pm 0.011$ (bootstrap error), which represents $\approx 5\sigma$ detection of positive $e$. This value is  statistically consistent with the median environmental gravitational field for these galaxies ($\langle e_{\rm env}\rangle = 0.034\pm 0.001$ (bootstrap error)). Furthermore, based on the robust binomial statistic with equal probabilities for $e>0$ and $e<0$, 78 cases of $e>0$ out of 113 is $4\sigma$ away from the expected mean of $56.5$ cases.

\begin{figure}
  \centering
  \includegraphics[width=1.\linewidth]{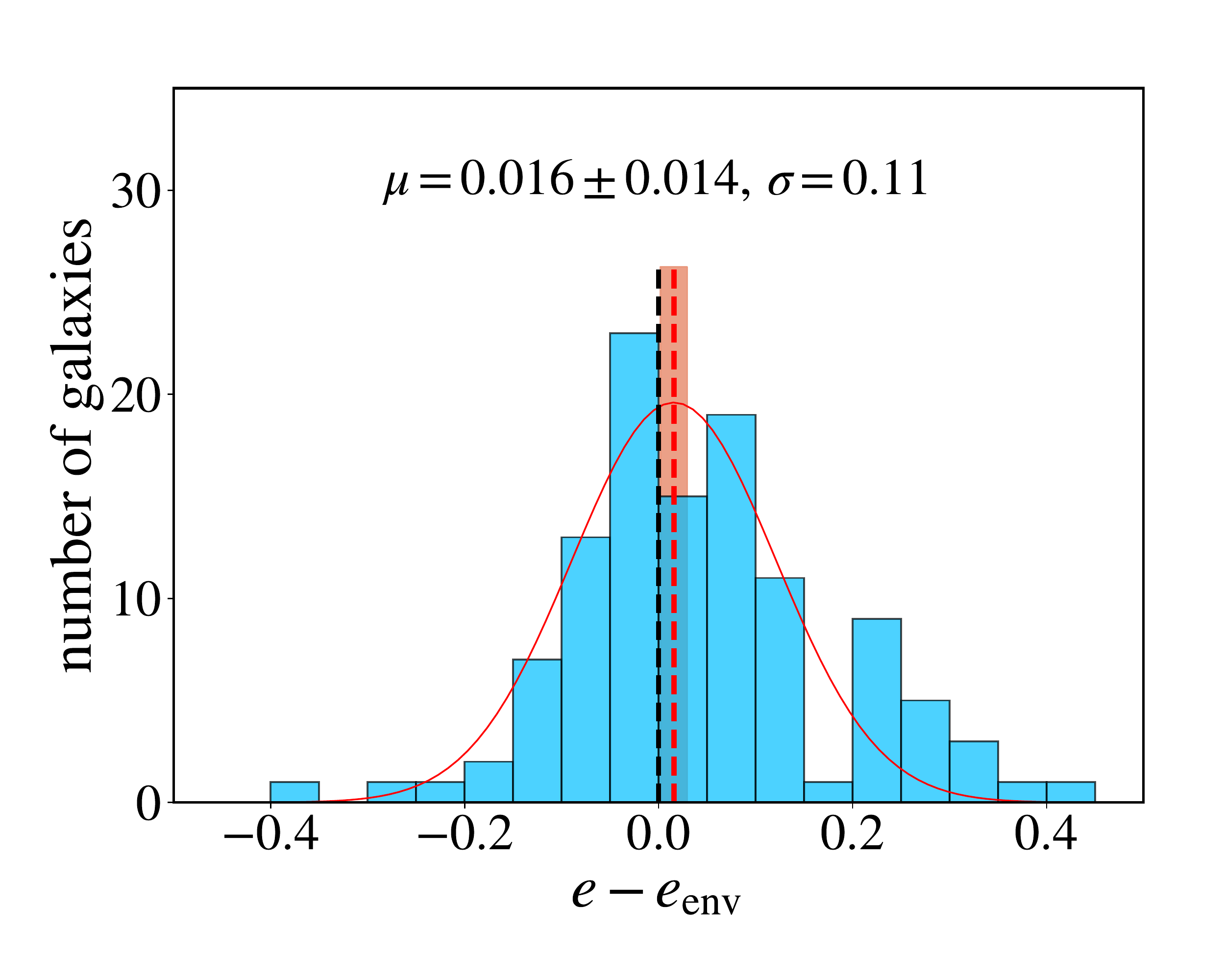}
    \vspace{-1.truecm}  
    \caption{\small 
    Comparison of external field strength estimates from kinematics versus analyzing the galaxies' environments.
 The distribution of $e$ derived from MCMC fits to the rotation curves is compared with that estimated from the observed environments of the galaxies ($e_{\rm env}$). There is good agreement up to the large uncertainties on the fitted values.
   }
   \label{dele}
\end{figure}

Fig.~\ref{dele} further shows the distribution of the individual difference $e-e_{\rm env}$. It has a broad distribution due to the large uncertainty in $e$ but is clearly consistent with zero: $\langle e-e_{\rm env} \rangle = 0.011 \pm 0.013$. It is intriguing that the mere fitting parameter $e$ returns, on average, the same value of the mean environmental gravitational field of the nearby Universe, computed in a fully independent way.

\subsubsection{Statistical properties of the posterior parameters of the galaxies} \label{sec:statprmt}

\begin{figure*}
  \centering
  \includegraphics[width=1.\linewidth]{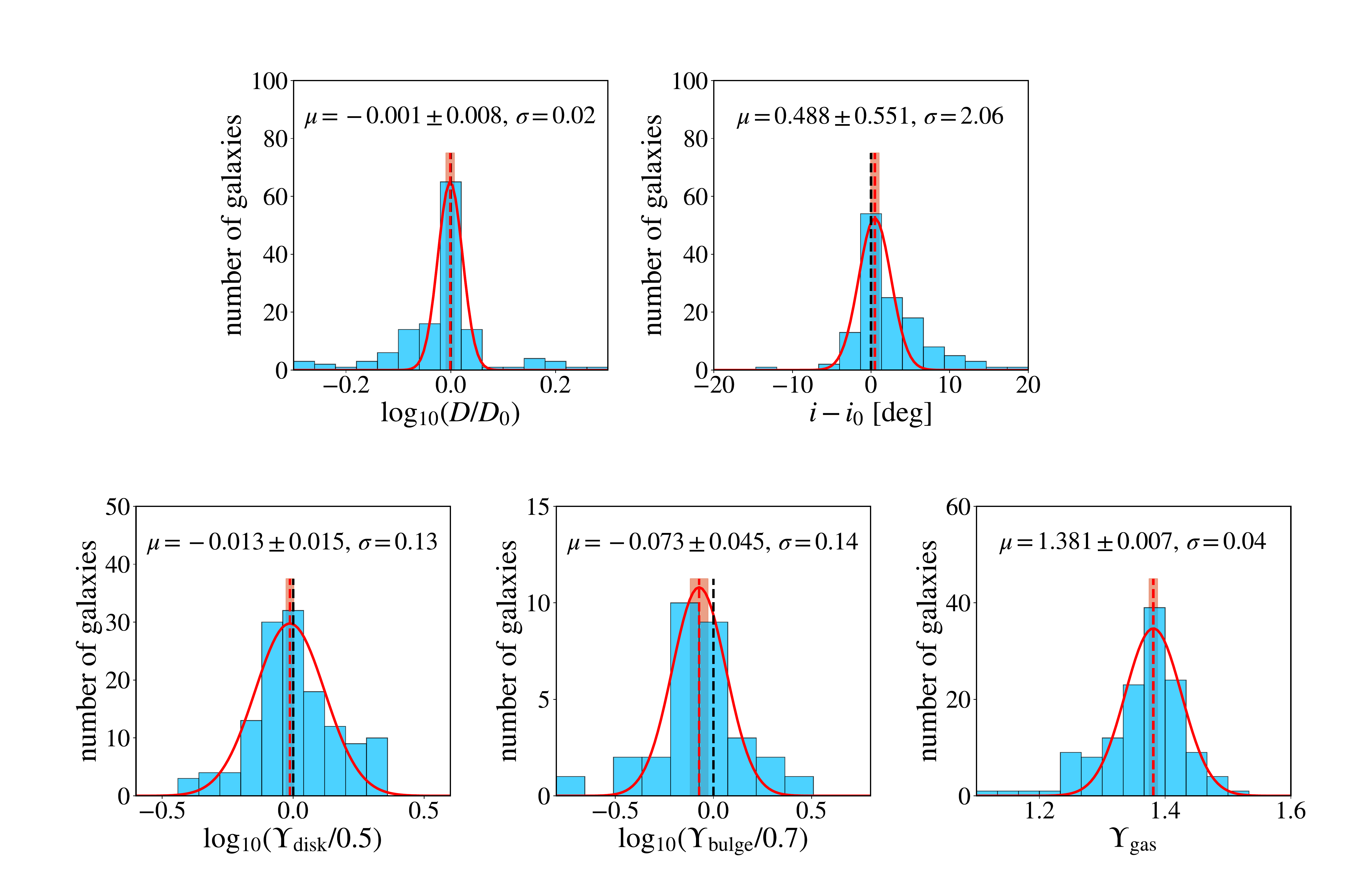}
    \vspace{-0.8truecm}  
    \caption{\small Fitted parameters for the 153 SPARC galaxies.
 The distributions of the fitted parameters from our MCMC simulations using Eq.~(\ref{eq:rare}) are compared with the SPARC measured or assumed values.
   }
   \label{prmtdist}
\end{figure*}

Fig.~\ref{prmtdist} shows the distribution of the parameters from the MCMC simulations with Eq.~(\ref{eq:rare}) for all 153 selected galaxies. The distribution of the distances is consistent with the SPARC reported values with an rms scatter of $0.02$~dex (5 percent). This is smaller than typical measurement uncertainties of $\sim 14$ percent. The posterior inclination angles are also consistent with the SPARC reported values with an rms scatter of $2.1^\circ$, smaller than typical measurement uncertainties of $\sim 4^\circ$. The distributions of the mass-to-light ratios ($\Upsilon_{\rm disk}$ and $\Upsilon_{\rm bul}$) for the disk and the bulge are consistent with the estimates from infrared studies, i.e., $\Upsilon_{\rm disk}=0.5$ $M_\odot /L_\odot$ and $\Upsilon_{\rm bul}=0.7$ $M_\odot /L_\odot$, with an rms scatter of $0.14$~dex. If anything $\Upsilon_{\rm bul}$ might be $0.6$, a little smaller than $0.7$. Finally, the distribution of $\Upsilon_{\rm gas}$ is in excellent agreement with $X^{-1}$ from Eq.~(\ref{eq:X1}), giving a mean value of $1.38\pm 0.04$ which is intermediate between a metal-poor dwarf galaxy with $X^{-1} = 1.34$ and a metal-rich giant spiral with $X^{-1} = 1.42$.

\subsubsection{Analysis of possible systematic effects} \label{sec:sys}

\begin{figure*}
  \centering
  \includegraphics[width=1.\linewidth]{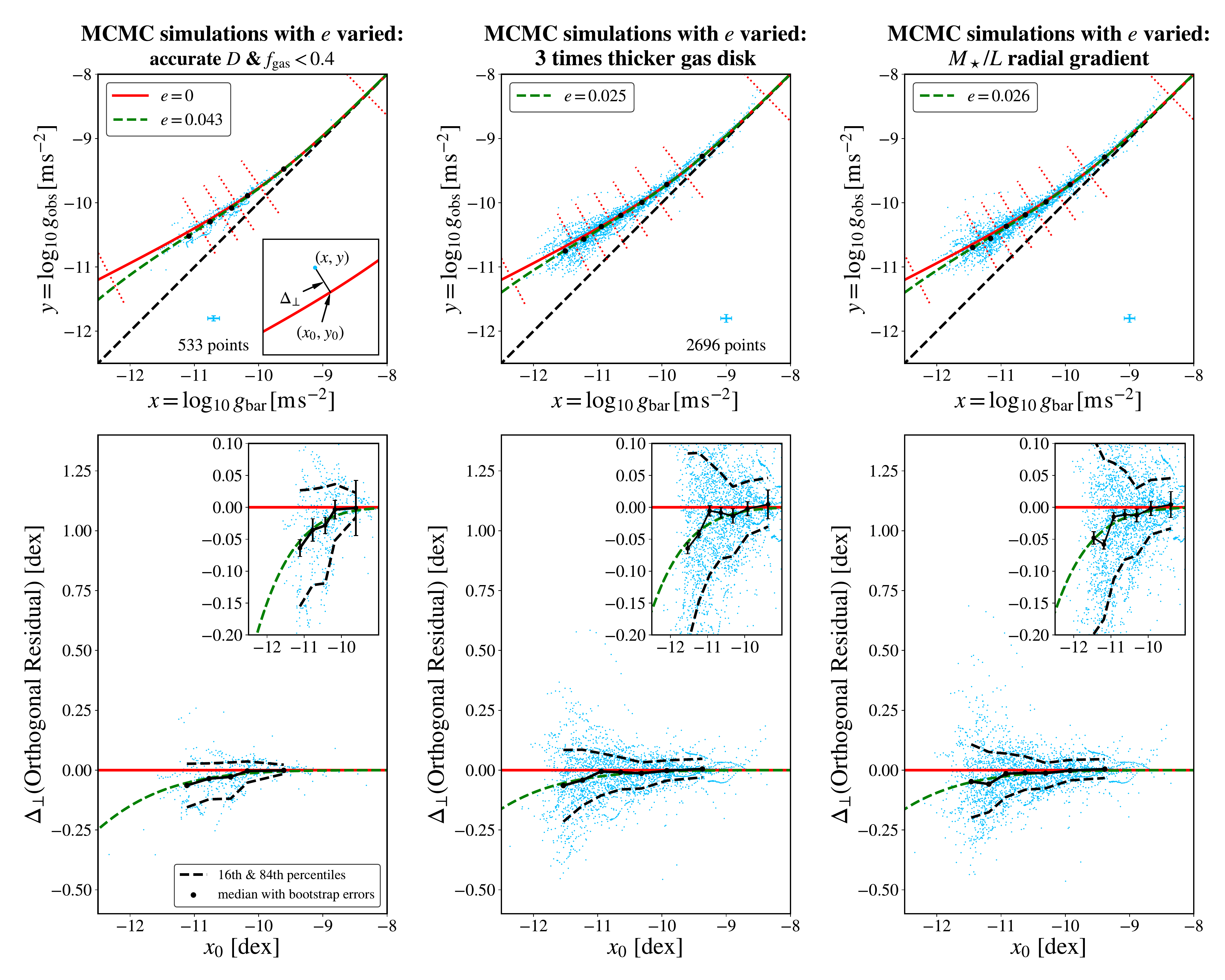}
    \vspace{-0.3truecm}  
    \caption{\small 
    Testing systematic uncertainties in the EFE detection.
 This figure has the same format as Fig.~\ref{rar} in the main paper. The left panels show a subset of data points with sub-dominant gas contribution $f_{\rm gas}<0.4$ and accurate distance measurements. The middle panels show the MCMC results assuming $3\times$ thicker gas disks for all galaxies. The right panels show the MCMC results assuming a radial $M_\star/L$ gradient in the stellar disks of all galaxies. Our conclusions on the EFE detection hold in all cases.
   }
   \label{rarsyst}
\end{figure*}

One may wonder whether the systematic deviations from Eq.~(\ref{eq:rar}) are due to some systematic uncertainties. There are three main observational effects that may systematically affect the low acceleration portion of the RAR: galaxy distances, the thickness of the gas disk, and possible variations of $M_\star/L$ in the stellar disk with radius. To mitigate the first two uncertainties, the left panel of Fig.~\ref{rarsyst} considers data points from galaxies with accurate distances based on the tip magnitude of the red giant branch, Cepheids, or Supernovae \citep{Lel16}, as well as low gas contributions ($f_{\rm gas}=M_{\rm gas}/M_{\rm bar} < 0.4$). Compared with Fig.~\ref{rar} in the main manuscript, it is clear that the scatter is smaller and the median trend is consistent with the full dataset.

The thickness of the gas disk is a concern because the EFE is detected in the galaxy outskirts, where the gas contribution becomes non-negligible or even dominating in some cases. Recent studies \citep{Bac19} suggest that gas disks may become thicker at large radii: this would systematically decrease $V_{\rm gas}$, hence $g_{\rm bar}$, moving points to the left of the RAR. Therefore, we repeat the MCMC fits considering gas disks that are three times thicker than assumed in the SPARC database. This is a very extreme scenario because not all galaxies will have such thick gas disks. Our goal is simply to provide an upper bound on the possible impact of this effect. Fig.~\ref{rarsyst} (middle panel) shows that there is still a significant systematic deviation from Eq.~(\ref{eq:rar}) even when we consider very thick gas disks.

Negative gradients of $M_\star/L$ with $R$ could also systematically decrease $V_{\rm disk}$, hence $g_{\rm bar}$, moving points to the left of the RAR. While we are treating the bulge separately in the most massive spirals (Sa to Sb), the stellar disk may potentially display a radial variation of its stellar populations. At 3.6 $\mu$m these variations have a relatively weak effect \citep{Schom19}, but we nevertheless repeat the MCMC fits considering a linear decrease in $\Upsilon_{\rm disk}$ by a factor of 2 from the center to the outermost observed radius. Again, this is an extreme scenario since most stellar disk are likely not showing such strong radial gradients in $\Upsilon_{\rm disk}$. Fig.~\ref{rarsyst} (right panel) shows that there is still a significant systematic deviation from Eq.~(\ref{eq:rar}).

\subsection{Comparison with previous results}
Only a few attempts have been made so far to detect the EFE from the RCs of galaxies \citep{Hag16,WK15}. In particular, \cite{Hag16} considered the RCs of 18 galaxies taken from the literature available at that time. These galaxies are known to have relatively nearby massive neighbors. Eleven of them are also included in our sample of 153 galaxies studied here. They are DDO 154, IC 2574, NGC 2998, NGC 3198, NGC 3521, NGC 3769, NGC 4100, NGC 4183, NGC 5033, NGC 5055, and NGC 5371.

\cite{Hag16} obtained values of $e$ ranging from about $0.1$ to $0.6$ with a median of $\sim 0.3$ and a typical uncertainty of $\sim 0.1$ for these 11 galaxies. Their values are systematically higher than our values ranging from about $ -0.1$ to $0.3$ with a median of $\sim 0.075$ and a typical uncertainty of $\sim 0.04$. This is primarily due to the fact that the disk models of \cite{Hag16} are based on a baryonic mass profile that declines more slowly than observed at large radii, requiring a larger EFE in the MOND context (a deficit of DM in the $\Lambda$CDM context).

There have also been indications of the EFE in pressure-supported galaxies \citep{MM13a,MM13b,FMM18,Krou18}. Pressure-supported galaxies are analyzed through their observed line-of-sight velocity dispersions. Because their stellar orbits are complex and not observed directly, a robust kinematic analysis to infer the EFE is challenging. However, \cite{MM13a, MM13b} have found that the observed velocity dispersions of the dwarf galaxies of the Andromeda galaxy are consistent with a MOND theory with EFE.  More recently, galaxies that appeared to have too low observed velocity dispersions and thus lack dark matter in the $\Lambda$CDM context \citep{vD18,vD19} may well be explained by the MOND EFE \citep{FMM18,Krou18,Mul19,Hag19}. 
  
\section{Discussion}  \label{sec:disc}

Galaxies of similar properties but subjected to different external gravitational fields show noticeably different rotation-curve behaviors at large radii (i.e.\ at very low accelerations).
Two galaxies in the strongest environmental fields show declining RCs in the outer parts, while two similar galaxies in the weakest environmental fields have flat RCs. 
The connection between internal dynamics and large-scale environment is corroborated by a statistical analysis of the entire SPARC sample. At accelerations 10 times lower than $g_\dagger$, the RAR is not fully described by a simple function of $g_{\rm bar}/g_\dagger$ (Eq.~\ref{eq:rar}) but requires an EFE-incorporated generalized function with an additional free parameter $e$ (Eq.~\ref{eq:nue}). Moreover, rotation-curve fits with Eq.~(\ref{eq:nue}) give a mean value of $e$ that is indistinguishable from the mean environmental gravitational field at the location of SPARC galaxies, computed in a fully independent fashion from the average distribution of mass in the nearby Universe. These results are summarized in Figs.~\ref{rar} and \ref{e_values}. Note that these results of fitting Eq.~(\ref{eq:nue}) to RCs are fully \emph{empirical}, independent of any theoretical interpretation. 

Can these results be explained in the standard $\Lambda$CDM framework? For the two golden massive galaxies subjected to strong large-scale gravitational field $g_\text{env}$, declining RCs are observed over a radial range of about 30 - 50 kpc, which are less than $\sim 15\%$ of the virial radius of the DM halo. 
Clearly, this is not the decline that should occur in the outer parts of $\Lambda$CDM halos, where the density profile decreases as $r^{-3}$, since we are probing the inner parts of the halo where the density profile goes approximately as $r^{-2}$, leading to flat RCs.

Thus, the only remaining option is represented by tidal forces. We calculated the expected tidal radii in $\Lambda$CDM using the formalism of \cite{King62}, taking the source of the tidal field to be the nearest 2M++ galaxy to the SPARC galaxy in question. We assume the source and test galaxies to have NFW \citep{NFW} halos following the $M_\star$-$M_\text{vir}$ relation of \cite{Krav18} and the $M_\text{vir}-$concentration relation of \cite{DK15}. We find the tidal radii to be much larger than the last measured points of the RCs, so the galaxies themselves are effectively shielded against large-scale tides.

The agreement between the MOND fitting parameter $e$ (Eq.~\ref{eq:rare}) and the environmental gravitational field $e_{\rm env}$ is an unpredicted result from the $\Lambda$CDM point of view. In principle, the baryon plus DM combination can combine to produce a declining rotation curve within tens of kpc as found here (i.e.\ $e>0$). For that matter, however, there is no a priori reason that the degree of declining must agree with the strength of the environmental gravitational field. There could have been an order-of-magnitude difference between $e$ and $e_{\rm env}$. Yet, we are seeing an interesting coincidence between the two.

Moreover, a downward deviation in the RAR near a tenth of $g_\dagger$ is not predicted by current $\Lambda$CDM state-of-the-art simulations or semi-analytical models \citep{DCL16,Des17,Nav17,KW17,Ten18} with some predicting the opposite trend \citep{Lud17,Fat18,Gar19}. To the best of our knowledge, there is no reported scenario in which the DM-baryon coupling in the outskirts of the disks depends on the external gravitational field from the large-scale galaxy environment in the manner found here. 

The empirical evidence is fully consistent with the EFE predicted by MOND modified gravity \citep{BM84}. More generally, our results suggest a violation of the SEP in rotationally-supported galaxies. While in GR the internal dynamics of a gravitationally-bound system is not affected by a uniform external field, our analysis indicates that external fields \emph{do} impact the internal dynamics. Our results are encouraging for modified gravity as an alternative (or modification) to the DM hypothesis and the standard $\Lambda$CDM cosmological model. They also highlight the path for future theoretical investigations of relativistic theories of gravity beyond GR (see, e.g., \citealt{Skor20}), possibly leading to a new cosmological model.

\section{Conclusions}  \label{sec:conc}

In this paper we provide observational evidence for the existence of the EFE (or a phenomenon akin to it) predicted by MOND modified gravity \citep{BM84}. We use accurate rotation curves and mass models from the SPARC database \citep{Lel16} and detect the EFE in three separate ways:
\begin{enumerate}

\item The EFE is individually detected in ``golden'' galaxies subjected to exceptionally strong external gravitational fields. The detection is highly significant ($11\sigma$ in NGC5055 and $8\sigma$ in NGC5033) and the best-fit values of the external gravitational fields are fully consistent with the independent estimates from the large-scale distribution of mass at the galaxies' location. Conversely, the EFE is not detected in control galaxies residing in the weakest external gravitational fields, as expected.

\item The EFE is statistically detected at more than 4$\sigma$ through a blind test using 153 SPARC galaxies. The mean value of the external gravitational field among the SPARC galaxies is again consistent with the independent estimate from the average distribution of mass in the nearby Universe.

\item The EFE also manifests as a small ($\ga 0.05$ dex), downward deviation from the empirical RAR occurring around $0.1 g_\dagger$. This behavior is not predicted by any of the existing galaxy-formation models in $\Lambda$CDM that were proposed to ``naturally'' reproduce the RAR. In contrast, this downward deviation \emph{is} predicted by the MOND modified gravity at the right acceleration scale.
  
\end{enumerate}

Our results suggest a breakdown of the SEP: the internal dynamics of a gravitational system in free-fall \textit{is} affected by a uniform external gravitational field. This sheds new light on the dark-matter problem and paves the way for relativistic theories of modified gravity in the weak-field regime of gravity $g \la 10^{-10}$~m~s$^{-2}$.

\vspace{0.3in}

\acknowledgements
We thank the organizers of the conference Bonn-Gravity 2019 (Pavel Kroupa and Indranil Banik) where several of these issues were brought to light. We thank Andrey Kravtsov for providing a code to calculate tidal radii of $\Lambda$CDM halos.
This work was supported by the National Research Foundation of Korea(NRF) grant funded by the Korea government(MSIT) (No.\ NRF-2019R1F1A1062477). HD is supported by St John's college, Oxford, and acknowledges financial support from ERC Grant No. 693024 and the Beecroft Trust. The Work of SSM is supported in part by NASA ADAP 80NSSC19k0570 and NSF PHY-1911909.
  



\newpage

\appendix

\section{The posterior PDFs of the parameters: golden and normal galaxies} 

\begin{figure*}[b]
  \centering
  \includegraphics[width=0.8\linewidth]{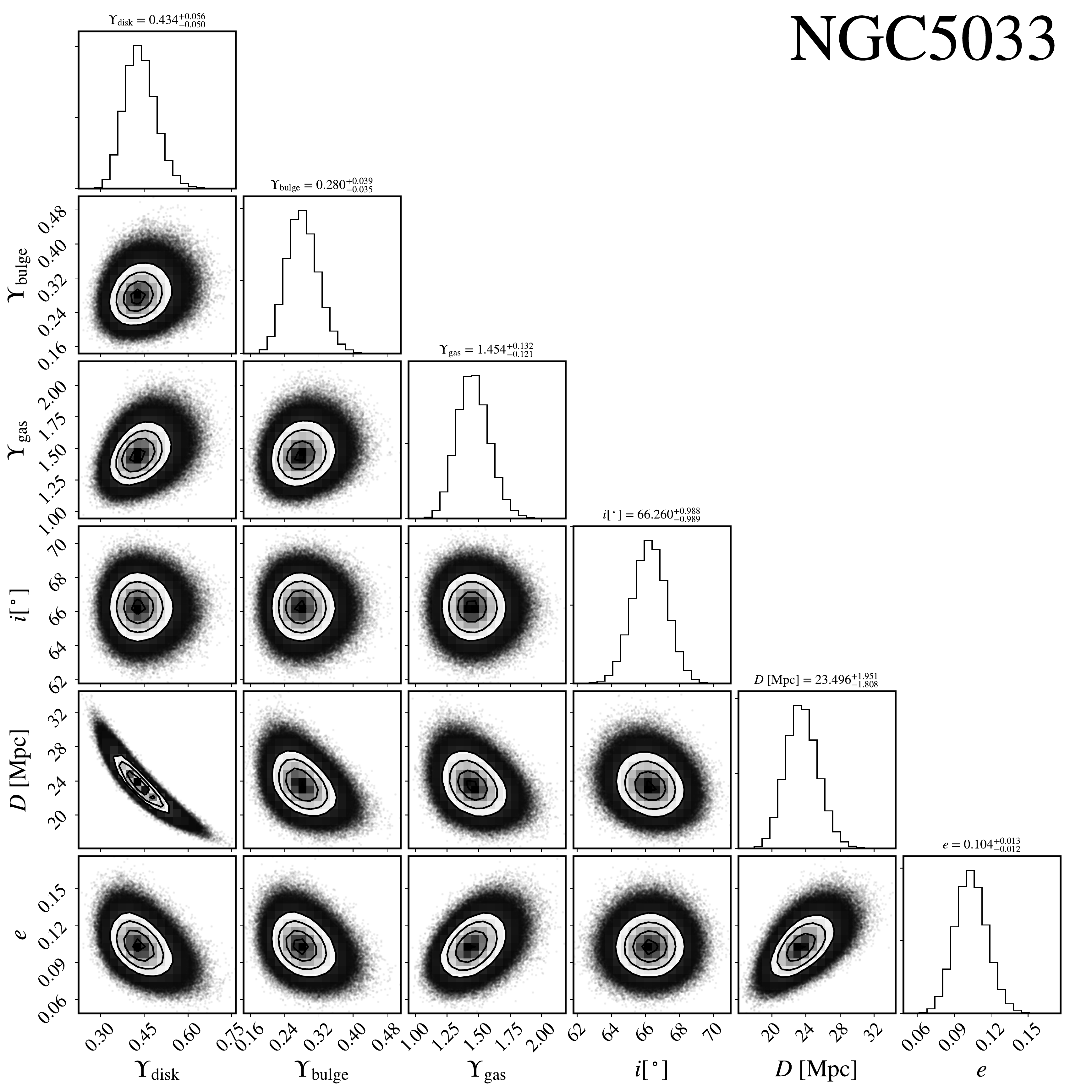}
    \vspace{-0.3truecm}  
    \caption{\small Parameter corner plot for NGC5033. 
 The posterior PDFs of the parameters for ``golden galaxy'' NGC5033 produced from MCMC simulations using Eq.~(\ref{eq:rare}).
   }
   \label{NGC5033}
    \vspace{-12pt}
\end{figure*}

\begin{figure*}
  \centering
  \includegraphics[width=0.8\linewidth]{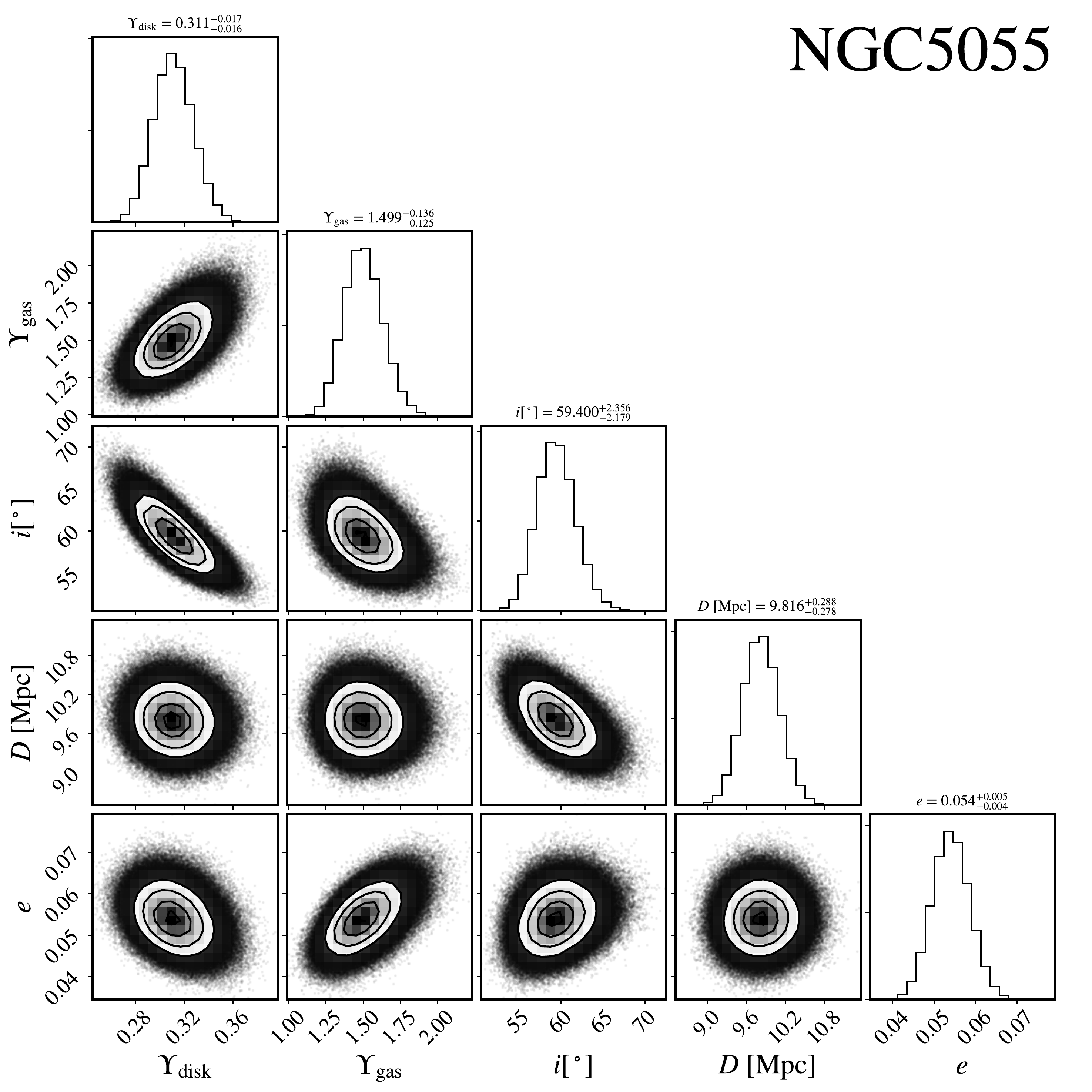}
    \vspace{-0.3truecm}  
    \caption{\small Parameter corner plot for NGC5055.
 The posterior PDFs of the parameters for ``golden galaxy'' NGC5055 produced from MCMC simulations using Eq.~(\ref{eq:rare}).
   }
   \label{NGC5055}
    \vspace{-12pt}
\end{figure*}

We present the full posterior PDFs of the parameters for the two golden galaxies NGC5033 and NGC5055, which are found in the strongest external fields among the SPARC galaxies. NGC5033 (Fig.~\ref{NGC5033}) has a bulge component while NGC5055 (Fig.~\ref{NGC5055}) does not. These are the cases in which $e$ is well constrained.

\begin{figure*}
  \centering
  \includegraphics[width=0.8\linewidth]{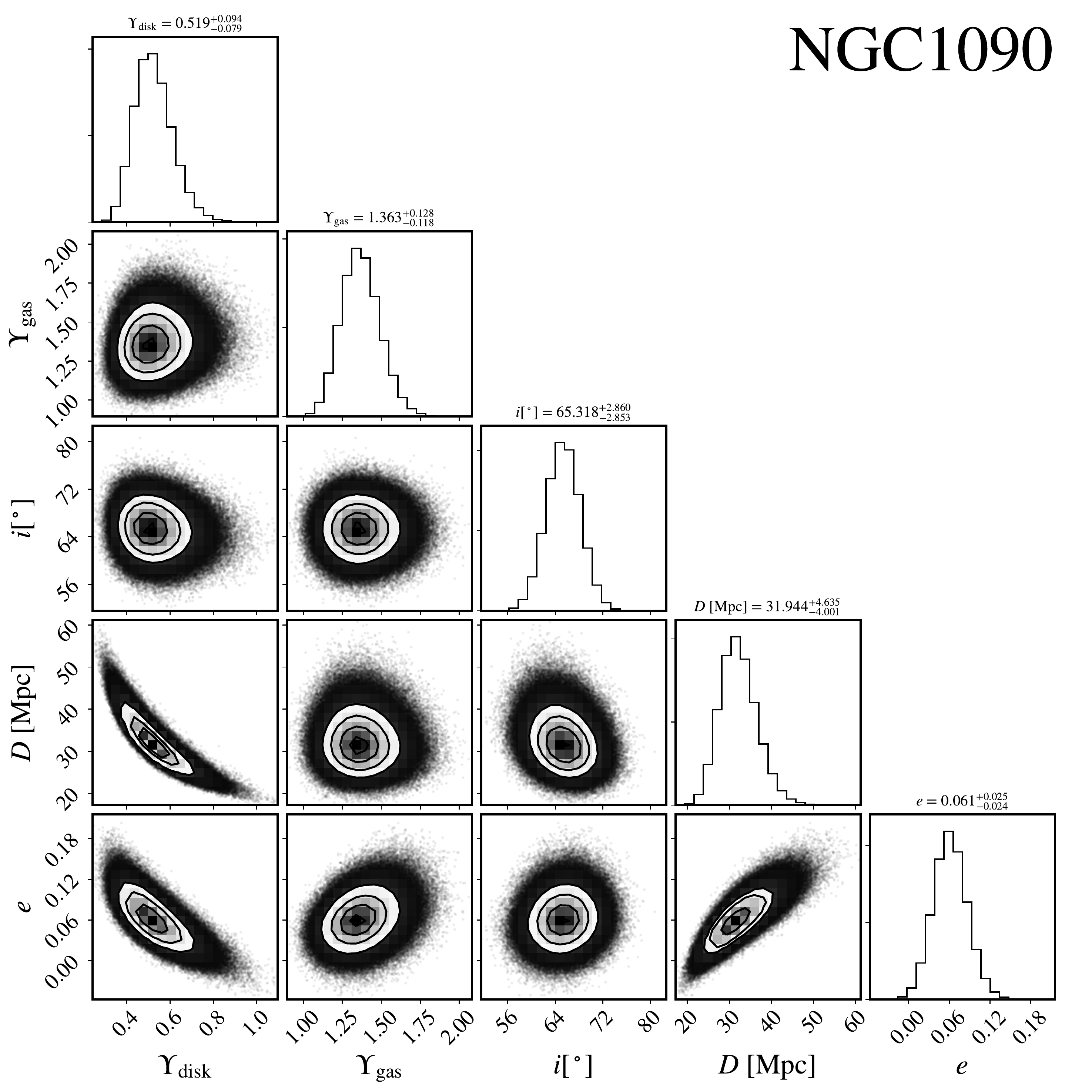}
    \vspace{-0.3truecm}  
    \caption{\small Parameter corner plot for NGC1090. 
 The posterior PDFs of the parameters for a control galaxy NGC1090 produced from MCMC simulations using Eq.~(\ref{eq:rare}).
   }
   \label{NGC1090}
    \vspace{-12pt}
\end{figure*}

\begin{figure*}
  \centering
  \includegraphics[width=0.8\linewidth]{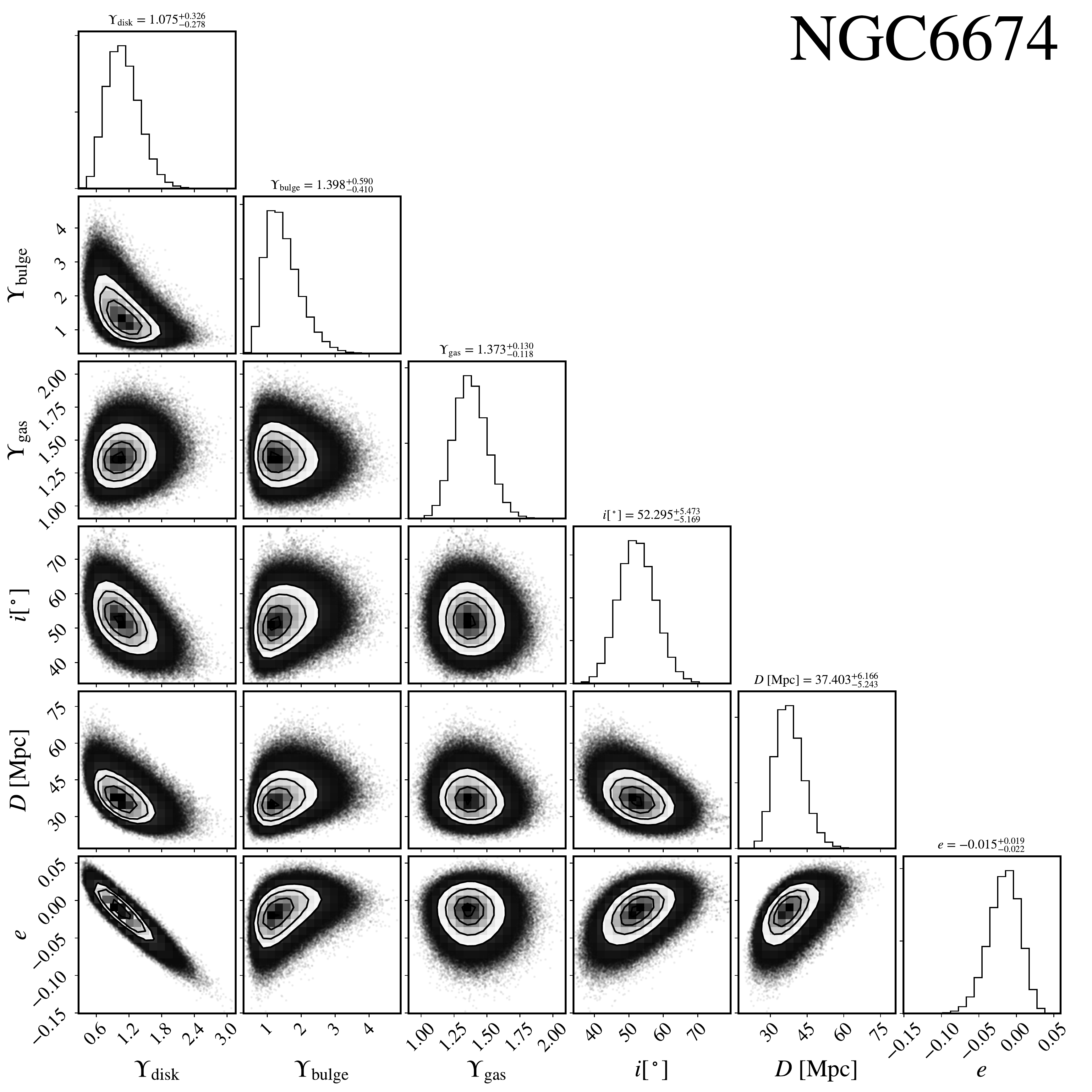}
    \vspace{-0.3truecm}  
    \caption{\small Parameter corner plot for NGC6674. 
 The posterior PDFs of the parameters for a control galaxy NGC6674 produced from MCMC simulations using Eq.~(\ref{eq:rare}).
   }
   \label{NGC6674}
    \vspace{-12pt}
\end{figure*}

\begin{figure*}
  \centering
  \includegraphics[width=0.8\linewidth]{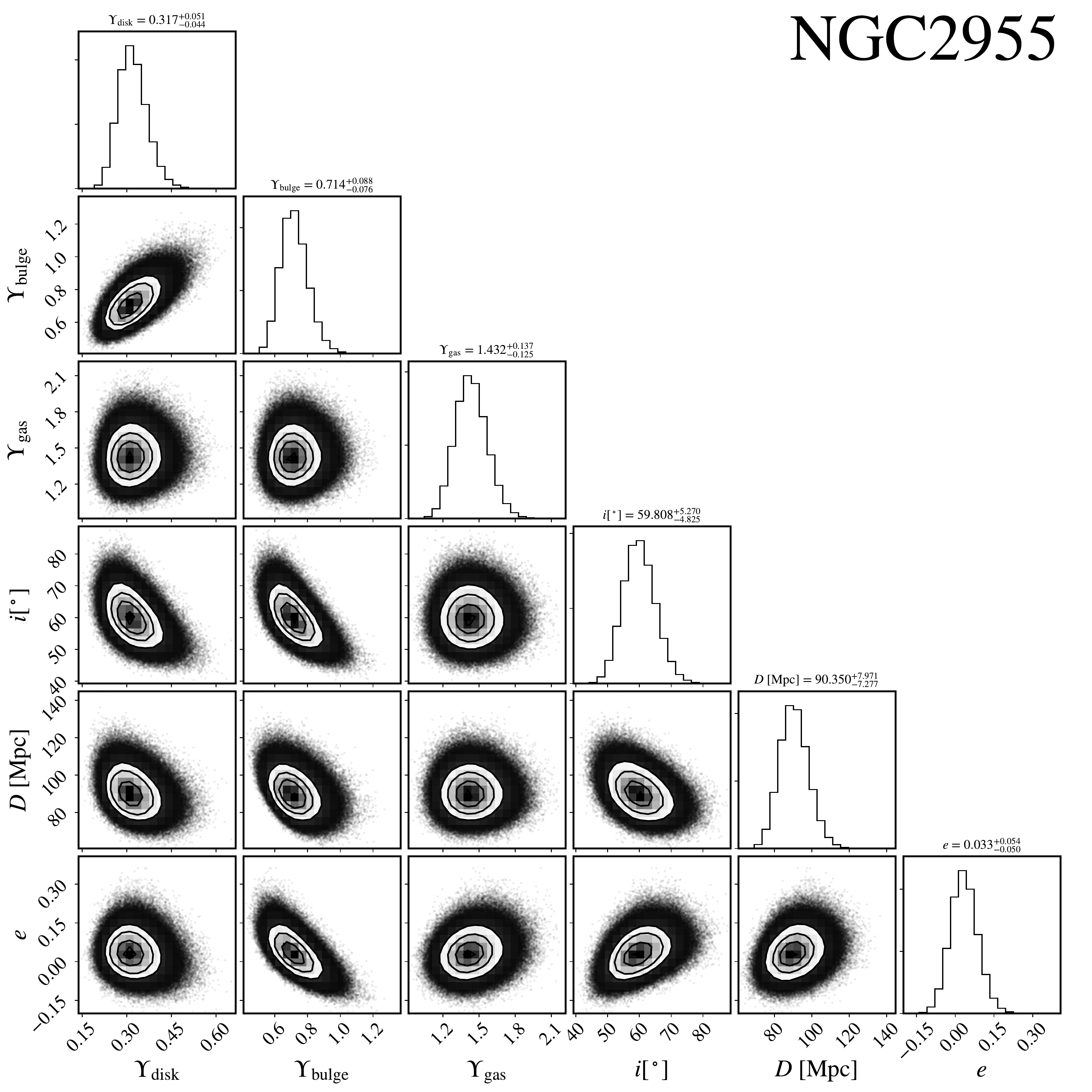}
    \vspace{-0.3truecm}  
    \caption{\small Parameter corner plot for NGC2955. 
 The posterior PDFs of the parameters for a normal galaxy NGC2955 produced from MCMC simulations using Eq.~(\ref{eq:rare}).
   }
   \label{NGC2955}
    \vspace{-12pt}
\end{figure*}

\begin{figure*}
  \centering
  \includegraphics[width=0.8\linewidth]{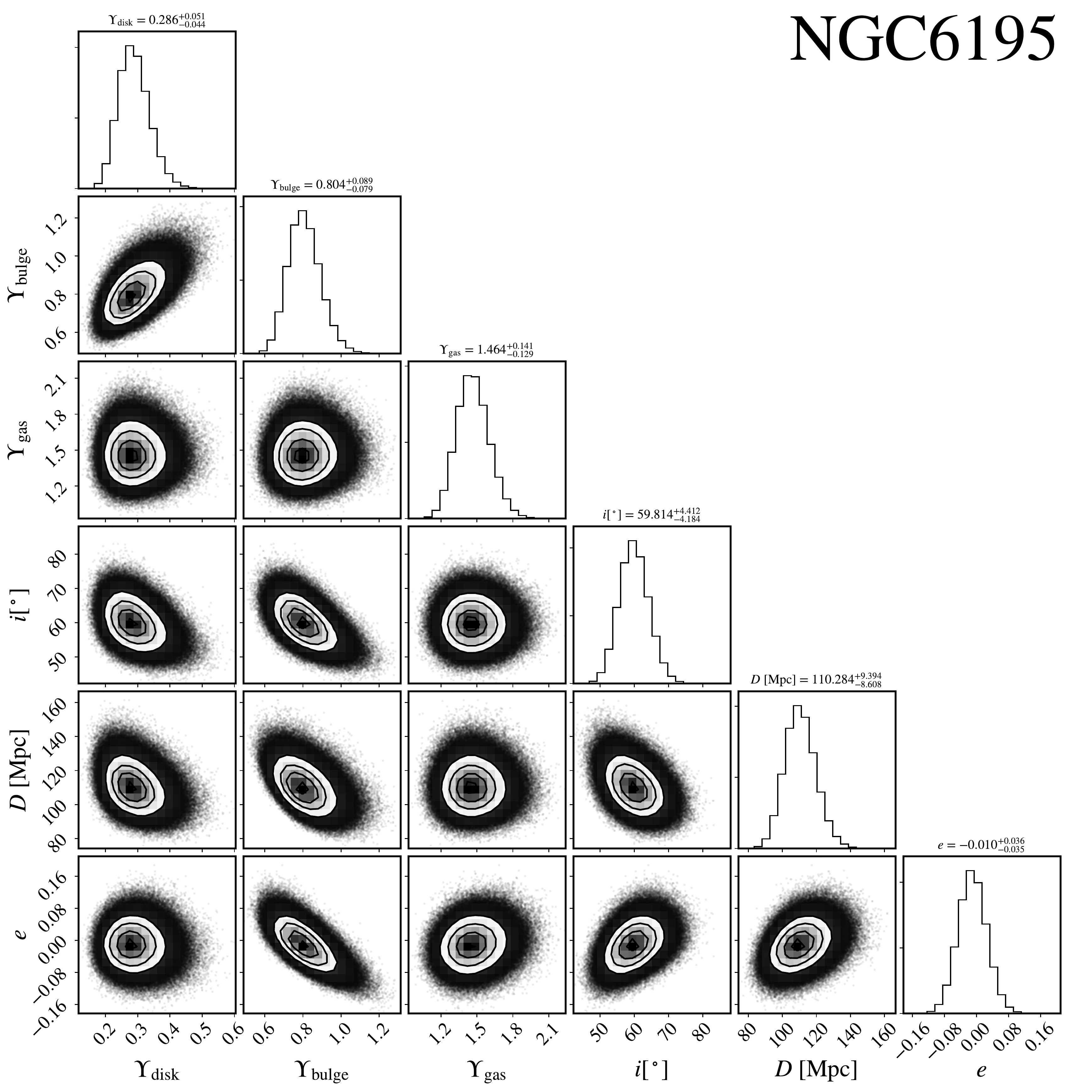}
    \vspace{-0.3truecm}  
    \caption{\small Parameter corner plot for NGC6195. 
 The posterior PDFs of the parameters for a normal galaxy NGC6195 produced from MCMC simulations using Eq.~(\ref{eq:rare}).
   }
   \label{NGC6195}
    \vspace{-12pt}
\end{figure*}

We also present the two control galaxies NGC1090 (Fig.~\ref{NGC1090}) and NGC6674 (Fig.~\ref{NGC6674}) that are found in the weakest external fields among the SPARC galaxies. These galaxies have statistical uncertainties of $\sim 0.02$ in $e$, which are lower than typical uncertainties of $\sim 0.04$ across the whole sample. Thus, we show another two examples, NGC2955 (Fig.~\ref{NGC2955}) and NGC6195 (Fig.~\ref{NGC6195}), that have statistical uncertainties of $\sim 0.04$ in $e$. Unlike the golden galaxies, $e$ is hardly constrained in these normal galaxies.

The corner plots for all 153 galaxies can be found at the webpage \mbox{http://astroweb.cwru.edu/SPARC/} and \mbox{http://home.sejong.ac.kr/$\sim$chae/}.

\section{Fitted values of the parameters}

The MCMC fitted values of the model parameters and the independent estimate of $e_{\rm env}$ from the environment can be found in the following table for the 153 SPARC galaxies. The MCMC value and its uncertainty come from the 50 percentile and the 15.9 and 84.1 percentiles of the posterior PDF. Note that the values of $e$ are not meaningful for galaxies with $\langle x_0 \rangle \ga -10$ because the EFE has little effects on the rotation velocites in the high acceleration range. See \S~\ref{sec:res_stat}.

\section*{Text of the Erratum}

While doing follow-up work, we noticed that the $e_{\rm env}$ values of six of the 175 SPARC galaxies were significantly ($>3\sigma$) overestimated due to an erroneous positioning of galaxies. The median of $e_{\rm env}$ of all SPARC galaxies remains unchanged, thus this error has no impact on the overall scientific contents or the conclusions of our paper. However, in Table~2 there was a mismatch between the original $e_{\rm env}$ values and the SPARC galaxies due to an indexing error in producing the table. Thus, we correct Table~2 with the right match and the corrected values. The table is also available at \mbox{http://astroweb.cwru.edu/SPARC/} and \mbox{http://home.sejong.ac.kr/$\sim$chae/}.

To account for the corrected $e_{\rm env}$ values for several galaxies, Figures 5 and 6 are revised. Our two ``golden galaxies'' NGC~5033 and NGC~5055 are no longer in exceptionally high-density environmental fields with $e_{\rm env}\sim 0.1$. However, they remain golden because they have high-quality and extended rotation curves allowing rare $>5 \sigma$ detection of $e$ and reside in environmental fields that are about five times stronger than those of the most isolated galaxies NGC 1090 and NGC 6674 of similar mass shown in Figure~2. Considering the corrected $e_{\rm env}$ values, the agreement between $e$ and $e_{\rm env}$ improves for NGC~5055 which has relatively more reliable data overall, reinforcing our conclusions.

\newpage

\newpage

\newpage

\newpage

\newpage

\newpage

\setcounter{table}{1}
\begin{table}
\caption{Fitted model parameters}
\begin{center}
  \begin{tabular}{lcccccccc} \hline
 galaxy  & $\langle x_0\rangle$   & $e$  & $e_{\rm env}$ & $D$ [Mpc]  & $i$ [$^\circ$] & $\Upsilon_{\rm gas}$ & $\Upsilon_{\rm disk}$  & $\Upsilon_{\rm bulge}$  \\
 \hline
       CamB & $ -11.736 $ & $ 0.461 _{ -0.054 } ^{ + 0.029 } $ & $ 0.030 _{ -0.002 } ^{ + 0.007 } $ & $ 3.07 _{ -0.21 } ^{ + 0.22 } $ & $ 59.04 _{ -4.63 } ^{ + 4.84 } $ & $ 1.27 _{ -0.11 } ^{ + 0.12 } $ & $ 0.32 _{ -0.05 } ^{ + 0.06 } $ & --- \\  
     D512-2 & $ -11.423 $ & $ 0.119 _{ -0.088 } ^{ + 0.136 } $ & $ 0.047 _{ -0.004 } ^{ + 0.008 } $ & $ 16.12 _{ -3.74 } ^{ + 4.76 } $ & $ 61.47 _{ -8.94 } ^{ + 9.02 } $ & $ 1.36 _{ -0.12 } ^{ + 0.13 } $ & $ 0.52 _{ -0.11 } ^{ + 0.13 } $ & --- \\  
     D564-8 & $ -11.987 $ & $ 0.067 _{ -0.027 } ^{ + 0.028 } $ & $ 0.031 _{ -0.003 } ^{ + 0.007 } $ & $ 8.75 _{ -0.27 } ^{ + 0.28 } $ & $ 61.53 _{ -7.60 } ^{ + 7.46 } $ & $ 1.37 _{ -0.12 } ^{ + 0.13 } $ & $ 0.41 _{ -0.08 } ^{ + 0.10 } $ & --- \\  
     D631-7 & $ -11.410 $ & $ -0.085 _{ -0.017 } ^{ + 0.016 } $ & $ 0.029 _{ -0.003 } ^{ + 0.005 } $ & $ 7.48 _{ -0.17 } ^{ + 0.17 } $ & $ 38.76 _{ -2.27 } ^{ + 2.35 } $ & $ 1.12 _{ -0.09 } ^{ + 0.10 } $ & $ 0.26 _{ -0.04 } ^{ + 0.05 } $ & --- \\  
     DDO064 & $ -11.080 $ & $ 0.063 _{ -0.084 } ^{ + 0.126 } $ & $ 0.034 _{ -0.005 } ^{ + 0.007 } $ & $ 6.97 _{ -1.55 } ^{ + 2.14 } $ & $ 63.29 _{ -4.65 } ^{ + 4.68 } $ & $ 1.35 _{ -0.12 } ^{ + 0.13 } $ & $ 0.53 _{ -0.10 } ^{ + 0.13 } $ & --- \\  
     DDO154 & $ -11.526 $ & $ 0.008 _{ -0.008 } ^{ + 0.008 } $ & $ 0.036 _{ -0.004 } ^{ + 0.007 } $ & $ 3.86 _{ -0.16 } ^{ + 0.17 } $ & $ 61.88 _{ -2.55 } ^{ + 2.65 } $ & $ 1.43 _{ -0.11 } ^{ + 0.12 } $ & $ 0.20 _{ -0.03 } ^{ + 0.03 } $ & --- \\  
     DDO161 & $ -11.281 $ & $ -0.044 _{ -0.010 } ^{ + 0.010 } $ & $ 0.043 _{ -0.006 } ^{ + 0.011 } $ & $ 3.37 _{ -0.30 } ^{ + 0.40 } $ & $ 73.91 _{ -7.97 } ^{ + 7.96 } $ & $ 1.37 _{ -0.12 } ^{ + 0.13 } $ & $ 0.33 _{ -0.06 } ^{ + 0.07 } $ & --- \\  
     DDO168 & $ -10.931 $ & $ -0.188 _{ -0.050 } ^{ + 0.043 } $ & $ 0.035 _{ -0.004 } ^{ + 0.006 } $ & $ 4.05 _{ -0.19 } ^{ + 0.20 } $ & $ 35.82 _{ -3.55 } ^{ + 3.72 } $ & $ 1.24 _{ -0.11 } ^{ + 0.12 } $ & $ 0.44 _{ -0.09 } ^{ + 0.11 } $ & --- \\  
     DDO170 & $ -11.461 $ & $ 0.038 _{ -0.018 } ^{ + 0.021 } $ & $ 0.057 _{ -0.009 } ^{ + 0.010 } $ & $ 11.82 _{ -1.51 } ^{ + 1.87 } $ & $ 66.33 _{ -6.40 } ^{ + 6.65 } $ & $ 1.28 _{ -0.11 } ^{ + 0.12 } $ & $ 0.67 _{ -0.12 } ^{ + 0.14 } $ & --- \\  
ESO079-G014 & $ -10.517 $ & $ -0.028 _{ -0.055 } ^{ + 0.058 } $ & $ 0.027 _{ -0.003 } ^{ + 0.002 } $ & $ 28.21 _{ -3.91 } ^{ + 4.57 } $ & $ 80.00 _{ -4.66 } ^{ + 4.56 } $ & $ 1.38 _{ -0.12 } ^{ + 0.13 } $ & $ 0.56 _{ -0.09 } ^{ + 0.11 } $ & --- \\  
ESO116-G012 & $ -10.715 $ & $ -0.062 _{ -0.040 } ^{ + 0.041 } $ & $ 0.026 _{ -0.004 } ^{ + 0.002 } $ & $ 13.50 _{ -2.03 } ^{ + 2.46 } $ & $ 74.69 _{ -2.92 } ^{ + 2.92 } $ & $ 1.39 _{ -0.12 } ^{ + 0.14 } $ & $ 0.45 _{ -0.08 } ^{ + 0.10 } $ & --- \\  
ESO444-G084 & $ -11.071 $ & $ -0.090 _{ -0.031 } ^{ + 0.027 } $ & $ 0.039 _{ -0.004 } ^{ + 0.016 } $ & $ 4.68 _{ -0.41 } ^{ + 0.45 } $ & $ 33.13 _{ -2.89 } ^{ + 3.03 } $ & $ 1.34 _{ -0.12 } ^{ + 0.13 } $ & $ 0.47 _{ -0.09 } ^{ + 0.12 } $ & --- \\  
ESO563-G021 & $ -10.059 $ & $ -0.031 _{ -0.033 } ^{ + 0.032 } $ & $ 0.035 _{ -0.006 } ^{ + 0.006 } $ & $ 66.48 _{ -7.21 } ^{ + 8.08 } $ & $ 83.63 _{ -2.81 } ^{ + 2.77 } $ & $ 1.45 _{ -0.13 } ^{ + 0.14 } $ & $ 0.69 _{ -0.09 } ^{ + 0.10 } $ & --- \\  
    F565-V2 & $ -11.436 $ & $ -0.013 _{ -0.047 } ^{ + 0.047 } $ & $ 0.024 _{ -0.002 } ^{ + 0.004 } $ & $ 52.32 _{ -8.90 } ^{ + 10.73 } $ & $ 65.91 _{ -8.91 } ^{ + 8.87 } $ & $ 1.36 _{ -0.12 } ^{ + 0.13 } $ & $ 0.50 _{ -0.10 } ^{ + 0.13 } $ & --- \\  
     F568-3 & $ -11.252 $ & $ 0.268 _{ -0.096 } ^{ + 0.117 } $ & $ 0.019 _{ -0.003 } ^{ + 0.001 } $ & $ 84.28 _{ -7.46 } ^{ + 8.09 } $ & $ 61.93 _{ -6.78 } ^{ + 6.95 } $ & $ 1.42 _{ -0.12 } ^{ + 0.13 } $ & $ 0.47 _{ -0.07 } ^{ + 0.08 } $ & --- \\  
    F568-V1 & $ -10.954 $ & $ 0.106 _{ -0.060 } ^{ + 0.077 } $ & $ 0.018 _{ -0.002 } ^{ + 0.003 } $ & $ 85.62 _{ -7.51 } ^{ + 8.25 } $ & $ 64.28 _{ -6.37 } ^{ + 6.70 } $ & $ 1.32 _{ -0.11 } ^{ + 0.13 } $ & $ 0.82 _{ -0.14 } ^{ + 0.16 } $ & --- \\  
     F571-8 & $ -10.730 $ & $ -0.349 _{ -0.041 } ^{ + 0.038 } $ & $ 0.030 _{ -0.004 } ^{ + 0.006 } $ & $ 27.74 _{ -3.51 } ^{ + 3.96 } $ & $ 82.84 _{ -5.16 } ^{ + 4.35 } $ & $ 1.38 _{ -0.12 } ^{ + 0.13 } $ & $ 0.23 _{ -0.04 } ^{ + 0.04 } $ & --- \\  
    F571-V1 & $ -11.479 $ & $ 0.229 _{ -0.121 } ^{ + 0.146 } $ & $ 0.019 _{ -0.002 } ^{ + 0.003 } $ & $ 80.12 _{ -7.16 } ^{ + 7.86 } $ & $ 44.11 _{ -7.86 } ^{ + 7.34 } $ & $ 1.38 _{ -0.12 } ^{ + 0.13 } $ & $ 0.47 _{ -0.09 } ^{ + 0.11 } $ & --- \\  
     F574-1 & $ -11.066 $ & $ 0.085 _{ -0.045 } ^{ + 0.054 } $ & $ 0.016 _{ -0.003 } ^{ + 0.004 } $ & $ 100.59 _{ -8.66 } ^{ + 9.53 } $ & $ 75.22 _{ -7.10 } ^{ + 7.04 } $ & $ 1.32 _{ -0.11 } ^{ + 0.12 } $ & $ 0.78 _{ -0.12 } ^{ + 0.14 } $ & --- \\  
     F583-1 & $ -11.337 $ & $ 0.035 _{ -0.052 } ^{ + 0.059 } $ & $ 0.033 _{ -0.004 } ^{ + 0.005 } $ & $ 36.15 _{ -6.89 } ^{ + 8.53 } $ & $ 67.66 _{ -4.48 } ^{ + 4.49 } $ & $ 1.24 _{ -0.10 } ^{ + 0.11 } $ & $ 0.96 _{ -0.14 } ^{ + 0.16 } $ & --- \\  
     F583-4 & $ -11.358 $ & $ 0.090 _{ -0.067 } ^{ + 0.092 } $ & $ 0.023 _{ -0.002 } ^{ + 0.004 } $ & $ 52.94 _{ -8.80 } ^{ + 10.66 } $ & $ 64.76 _{ -8.19 } ^{ + 8.35 } $ & $ 1.37 _{ -0.12 } ^{ + 0.13 } $ & $ 0.49 _{ -0.10 } ^{ + 0.12 } $ & --- \\  
     IC2574 & $ -11.722 $ & $ 0.075 _{ -0.015 } ^{ + 0.017 } $ & $ 0.029 _{ -0.003 } ^{ + 0.004 } $ & $ 4.01 _{ -0.18 } ^{ + 0.19 } $ & $ 81.45 _{ -4.84 } ^{ + 4.57 } $ & $ 1.66 _{ -0.12 } ^{ + 0.13 } $ & $ 0.20 _{ -0.03 } ^{ + 0.03 } $ & --- \\  
     IC4202 & $ -9.998  $ & $ 0.187 _{ -0.056 } ^{ + 0.064 } $ & $ 0.033 _{ -0.009 } ^{ + 0.008 } $ & $ 99.16 _{ -7.23 } ^{ + 7.71 } $ & $ 89.33 _{ -0.73 } ^{ + 0.47 } $ & $ 1.26 _{ -0.11 } ^{ + 0.11 } $ & $ 0.90 _{ -0.09 } ^{ + 0.09 } $ & $ 0.44 _{ -0.04 } ^{ + 0.04 } $  \\  
   KK98-251 & $ -11.423 $ & $ 0.293 _{ -0.111 } ^{ + 0.125 } $ & $ 0.031 _{ -0.001 } ^{ + 0.006 } $ & $ 7.21 _{ -1.39 } ^{ + 1.42 } $ & $ 62.71 _{ -4.50 } ^{ + 4.54 } $ & $ 1.37 _{ -0.12 } ^{ + 0.13 } $ & $ 0.47 _{ -0.09 } ^{ + 0.12 } $ & --- \\  
    NGC0024 & $ -10.096 $ & $ -0.004 _{ -0.016 } ^{ + 0.017 } $ & $ 0.028 _{ -0.003 } ^{ + 0.003 } $ & $ 7.48 _{ -0.34 } ^{ + 0.35 } $ & $ 67.27 _{ -2.68 } ^{ + 2.71 } $ & $ 1.34 _{ -0.12 } ^{ + 0.13 } $ & $ 0.99 _{ -0.10 } ^{ + 0.11 } $ & --- \\  
    NGC0055 & $ -11.110 $ & $ 0.052 _{ -0.020 } ^{ + 0.023 } $ & $ 0.032 _{ -0.005 } ^{ + 0.006 } $ & $ 1.94 _{ -0.09 } ^{ + 0.09 } $ & $ 75.23 _{ -3.11 } ^{ + 3.14 } $ & $ 1.33 _{ -0.11 } ^{ + 0.12 } $ & $ 0.21 _{ -0.03 } ^{ + 0.04 } $ & --- \\  
    NGC0100 & $ -10.989 $ & $ -0.099 _{ -0.038 } ^{ + 0.035 } $ & $ 0.024 _{ -0.003 } ^{ + 0.001 } $ & $ 9.72 _{ -1.56 } ^{ + 1.89 } $ & $ 88.80 _{ -0.91 } ^{ + 0.74 } $ & $ 1.39 _{ -0.12 } ^{ + 0.14 } $ & $ 0.39 _{ -0.07 } ^{ + 0.09 } $ & --- \\  
    NGC0247 & $ -10.915 $ & $ 0.202 _{ -0.055 } ^{ + 0.069 } $ & $ 0.028 _{ -0.002 } ^{ + 0.006 } $ & $ 3.76 _{ -0.18 } ^{ + 0.19 } $ & $ 75.59 _{ -2.83 } ^{ + 2.83 } $ & $ 1.29 _{ -0.11 } ^{ + 0.12 } $ & $ 1.03 _{ -0.11 } ^{ + 0.12 } $ & --- \\  
    NGC0289 & $ -11.225 $ & $ 0.125 _{ -0.029 } ^{ + 0.036 } $ & $ 0.024 _{ -0.002 } ^{ + 0.004 } $ & $ 19.93 _{ -2.65 } ^{ + 3.17 } $ & $ 54.37 _{ -3.99 } ^{ + 4.10 } $ & $ 1.43 _{ -0.12 } ^{ + 0.13 } $ & $ 0.44 _{ -0.06 } ^{ + 0.07 } $ & --- \\  
    NGC0300 & $ -11.083 $ & $ -0.009 _{ -0.025 } ^{ + 0.026 } $ & $ 0.029 _{ -0.002 } ^{ + 0.007 } $ & $ 2.03 _{ -0.09 } ^{ + 0.10 } $ & $ 47.31 _{ -4.58 } ^{ + 5.37 } $ & $ 1.34 _{ -0.12 } ^{ + 0.13 } $ & $ 0.40 _{ -0.06 } ^{ + 0.08 } $ & --- \\  
    NGC0801 & $ -10.204 $ & $ 0.190 _{ -0.027 } ^{ + 0.029 } $ & $ 0.042 _{ -0.012 } ^{ + 0.007 } $ & $ 68.49 _{ -5.86 } ^{ + 6.51 } $ & $ 79.93 _{ -1.00 } ^{ + 1.00 } $ & $ 1.44 _{ -0.13 } ^{ + 0.14 } $ & $ 0.60 _{ -0.06 } ^{ + 0.07 } $ & --- \\  
    NGC0891 & $ -10.025  $ & $ -0.110 _{ -0.019 } ^{ + 0.020 } $ & $ 0.025 _{ -0.003 } ^{ + 0.002 } $ & $ 9.84 _{ -0.44 } ^{ + 0.46 } $ & $ 89.33 _{ -0.73 } ^{ + 0.47 } $ & $ 1.34 _{ -0.11 } ^{ + 0.13 } $ & $ 0.33 _{ -0.02 } ^{ + 0.02 } $ & $ 0.52 _{ -0.06 } ^{ + 0.06 } $  \\  
    NGC1003 & $ -11.250 $ & $ -0.054 _{ -0.008 } ^{ + 0.008 } $ & $ 0.023 _{ -0.003 } ^{ + 0.002 } $ & $ 6.54 _{ -0.58 } ^{ + 0.65 } $ & $ 70.18 _{ -4.48 } ^{ + 4.54 } $ & $ 1.22 _{ -0.10 } ^{ + 0.11 } $ & $ 0.77 _{ -0.10 } ^{ + 0.11 } $ & --- \\  
    NGC1090 & $ -10.626 $ & $ 0.061 _{ -0.024 } ^{ + 0.025 } $ & $ 0.011 _{ -0.001 } ^{ + 0.002 } $ & $ 31.94 _{ -4.02 } ^{ + 4.66 } $ & $ 65.32 _{ -2.87 } ^{ + 2.88 } $ & $ 1.36 _{ -0.12 } ^{ + 0.13 } $ & $ 0.52 _{ -0.08 } ^{ + 0.09 } $ & --- \\  
    NGC2403 & $ -10.435 $ & $ -0.019 _{ -0.005 } ^{ + 0.005 } $ & $ 0.031 _{ -0.002 } ^{ + 0.005 } $ & $ 3.59 _{ -0.13 } ^{ + 0.13 } $ & $ 72.06 _{ -2.21 } ^{ + 2.26 } $ & $ 0.76 _{ -0.05 } ^{ + 0.06 } $ & $ 0.39 _{ -0.02 } ^{ + 0.02 } $ & --- \\  
    NGC2683 & $ -10.612  $ & $ 0.091 _{ -0.028 } ^{ + 0.031 } $ & $ 0.033 _{ -0.004 } ^{ + 0.006 } $ & $ 9.88 _{ -0.45 } ^{ + 0.47 } $ & $ 81.01 _{ -4.53 } ^{ + 4.37 } $ & $ 1.41 _{ -0.12 } ^{ + 0.14 } $ & $ 0.56 _{ -0.05 } ^{ + 0.05 } $ & $ 0.69 _{ -0.14 } ^{ + 0.17 } $  \\  
    NGC2841 & $ -9.797  $ & $ -0.027 _{ -0.013 } ^{ + 0.012 } $ & $ 0.037 _{ -0.004 } ^{ + 0.002 } $ & $ 14.03 _{ -0.91 } ^{ + 0.98 } $ & $ 82.99 _{ -5.41 } ^{ + 4.43 } $ & $ 1.31 _{ -0.11 } ^{ + 0.12 } $ & $ 0.91 _{ -0.09 } ^{ + 0.10 } $ & $ 0.96 _{ -0.07 } ^{ + 0.08 } $  \\  
    NGC2903 & $ -10.616 $ & $ 0.040 _{ -0.008 } ^{ + 0.008 } $ & $ 0.031 _{ -0.002 } ^{ + 0.006 } $ & $ 12.47 _{ -0.89 } ^{ + 0.97 } $ & $ 69.15 _{ -2.75 } ^{ + 2.76 } $ & $ 1.26 _{ -0.10 } ^{ + 0.11 } $ & $ 0.18 _{ -0.02 } ^{ + 0.02 } $ & --- \\  
    NGC2915 & $ -11.531 $ & $ -0.052 _{ -0.013 } ^{ + 0.012 } $ & $ 0.035 _{ -0.006 } ^{ + 0.008 } $ & $ 4.12 _{ -0.19 } ^{ + 0.20 } $ & $ 62.30 _{ -3.37 } ^{ + 3.41 } $ & $ 1.35 _{ -0.12 } ^{ + 0.13 } $ & $ 0.58 _{ -0.09 } ^{ + 0.11 } $ & --- \\  
    NGC2955 & $ -9.783  $ & $ 0.033 _{ -0.051 } ^{ + 0.054 } $ & $ 0.017 _{ -0.004 } ^{ + 0.001 } $ & $ 90.35 _{ -7.31 } ^{ + 8.02 } $ & $ 59.81 _{ -4.85 } ^{ + 5.30 } $ & $ 1.43 _{ -0.13 } ^{ + 0.14 } $ & $ 0.32 _{ -0.04 } ^{ + 0.05 } $ & $ 0.71 _{ -0.08 } ^{ + 0.09 } $  \\  
    NGC2976 & $ -10.365 $ & $ 0.387 _{ -0.114 } ^{ + 0.080 } $ & $ 0.032 _{ -0.003 } ^{ + 0.003 } $ & $ 3.62 _{ -0.17 } ^{ + 0.17 } $ & $ 76.32 _{ -6.13 } ^{ + 6.27 } $ & $ 1.44 _{ -0.12 } ^{ + 0.13 } $ & $ 0.44 _{ -0.05 } ^{ + 0.05 } $ & --- \\  
    NGC2998 & $ -10.488 $ & $ 0.110 _{ -0.030 } ^{ + 0.033 } $ & $ 0.027 _{ -0.006 } ^{ + 0.007 } $ & $ 70.00 _{ -7.59 } ^{ + 8.65 } $ & $ 58.69 _{ -1.94 } ^{ + 1.94 } $ & $ 1.45 _{ -0.13 } ^{ + 0.14 } $ & $ 0.54 _{ -0.07 } ^{ + 0.09 } $ & --- \\  
    NGC3109 & $ -11.513 $ & $ 0.012 _{ -0.010 } ^{ + 0.010 } $ & $ 0.031 _{ -0.003 } ^{ + 0.007 } $ & $ 1.40 _{ -0.06 } ^{ + 0.07 } $ & $ 76.86 _{ -3.81 } ^{ + 3.91 } $ & $ 1.68 _{ -0.13 } ^{ + 0.14 } $ & $ 0.24 _{ -0.04 } ^{ + 0.05 } $ & --- \\  
    NGC3198 & $ -10.613 $ & $ 0.057 _{ -0.012 } ^{ + 0.013 } $ & $ 0.036 _{ -0.002 } ^{ + 0.006 } $ & $ 15.28 _{ -1.08 } ^{ + 1.17 } $ & $ 75.62 _{ -2.69 } ^{ + 2.72 } $ & $ 1.36 _{ -0.12 } ^{ + 0.13 } $ & $ 0.43 _{ -0.04 } ^{ + 0.04 } $ & --- \\  
    NGC3521 & $ -9.407 $ & $ -0.106 _{ -0.057 } ^{ + 0.055 } $ & $ 0.033 _{ -0.003 } ^{ + 0.011 } $ & $ 6.65 _{ -0.96 } ^{ + 1.12 } $ & $ 78.31 _{ -4.35 } ^{ + 4.34 } $ & $ 1.42 _{ -0.13 } ^{ + 0.14 } $ & $ 0.56 _{ -0.09 } ^{ + 0.10 } $ & --- \\  
 \hline
\end{tabular}
\end{center}
\end{table}

\newpage

\setcounter{table}{1}
\begin{table}
\caption{(continued) Fitted model parameters}
\begin{center}
  \begin{tabular}{lcccccccc} \hline
    galaxy  & $\langle x_0\rangle$   & $e$  & $e_{\rm env}$ & $D$ [Mpc] & $i$ [$^\circ$] & $\Upsilon_{\rm gas}$ & $\Upsilon_{\rm disk}$  & $\Upsilon_{\rm bulge}$  \\
    \hline
    NGC3726 & $ -10.485 $ & $ -0.001 _{ -0.030 } ^{ + 0.032 } $ & $ 0.038 _{ -0.005 } ^{ + 0.004 } $ & $ 14.36 _{ -1.39 } ^{ + 1.54 } $ & $ 52.20 _{ -1.95 } ^{ + 1.96 } $ & $ 1.34 _{ -0.12 } ^{ + 0.13 } $ & $ 0.44 _{ -0.06 } ^{ + 0.07 } $ & --- \\  
    NGC3741 & $ -11.765 $ & $ -0.015 _{ -0.009 } ^{ + 0.009 } $ & $ 0.032 _{ -0.003 } ^{ + 0.008 } $ & $ 3.10 _{ -0.15 } ^{ + 0.16 } $ & $ 69.55 _{ -3.86 } ^{ + 3.93 } $ & $ 1.35 _{ -0.12 } ^{ + 0.13 } $ & $ 0.34 _{ -0.06 } ^{ + 0.06 } $ & --- \\  
    NGC3769 & $ -10.927 $ & $ 0.022 _{ -0.018 } ^{ + 0.019 } $ & $ 0.037 _{ -0.003 } ^{ + 0.003 } $ & $ 17.36 _{ -1.59 } ^{ + 1.75 } $ & $ 70.20 _{ -1.97 } ^{ + 1.97 } $ & $ 1.41 _{ -0.12 } ^{ + 0.14 } $ & $ 0.38 _{ -0.06 } ^{ + 0.07 } $ & --- \\  
    NGC3877 & $ -10.082 $ & $ 0.253 _{ -0.163 } ^{ + 0.158 } $ & $ 0.038 _{ -0.005 } ^{ + 0.005 } $ & $ 17.39 _{ -1.81 } ^{ + 2.01 } $ & $ 76.04 _{ -1.00 } ^{ + 1.00 } $ & $ 1.39 _{ -0.12 } ^{ + 0.13 } $ & $ 0.50 _{ -0.07 } ^{ + 0.08 } $ & --- \\  
    NGC3893 & $ -10.297 $ & $ -0.029 _{ -0.046 } ^{ + 0.047 } $ & $ 0.037 _{ -0.005 } ^{ + 0.005 } $ & $ 18.33 _{ -1.86 } ^{ + 2.07 } $ & $ 49.72 _{ -1.91 } ^{ + 1.91 } $ & $ 1.41 _{ -0.12 } ^{ + 0.14 } $ & $ 0.45 _{ -0.06 } ^{ + 0.07 } $ & --- \\  
    NGC3917 & $ -10.601 $ & $ 0.121 _{ -0.055 } ^{ + 0.067 } $ & $ 0.040 _{ -0.005 } ^{ + 0.012 } $ & $ 19.41 _{ -2.05 } ^{ + 2.32 } $ & $ 79.34 _{ -1.96 } ^{ + 1.97 } $ & $ 1.39 _{ -0.12 } ^{ + 0.13 } $ & $ 0.61 _{ -0.08 } ^{ + 0.10 } $ & --- \\  
    NGC3949 & $ -9.859 $ & $ 0.005 _{ -0.168 } ^{ + 0.228 } $ & $ 0.036 _{ -0.003 } ^{ + 0.008 } $ & $ 17.03 _{ -1.86 } ^{ + 2.08 } $ & $ 55.03 _{ -1.97 } ^{ + 1.97 } $ & $ 1.40 _{ -0.12 } ^{ + 0.13 } $ & $ 0.43 _{ -0.06 } ^{ + 0.07 } $ & --- \\  
    NGC3953 & $ -9.950 $ & $ 0.393 _{ -0.120 } ^{ + 0.077 } $ & $ 0.040 _{ -0.004 } ^{ + 0.005 } $ & $ 18.62 _{ -1.93 } ^{ + 2.15 } $ & $ 62.11 _{ -0.99 } ^{ + 0.99 } $ & $ 1.41 _{ -0.12 } ^{ + 0.14 } $ & $ 0.59 _{ -0.07 } ^{ + 0.08 } $ & --- \\  
    NGC3972 & $ -10.494 $ & $ -0.087 _{ -0.063 } ^{ + 0.072 } $ & $ 0.047 _{ -0.006 } ^{ + 0.003 } $ & $ 17.15 _{ -1.93 } ^{ + 2.18 } $ & $ 77.02 _{ -1.00 } ^{ + 1.00 } $ & $ 1.38 _{ -0.12 } ^{ + 0.13 } $ & $ 0.46 _{ -0.08 } ^{ + 0.09 } $ & --- \\  
    NGC3992 & $ -10.447 $ & $ 0.101 _{ -0.031 } ^{ + 0.034 } $ & $ 0.033 _{ -0.003 } ^{ + 0.007 } $ & $ 24.46 _{ -1.93 } ^{ + 2.10 } $ & $ 56.76 _{ -1.92 } ^{ + 1.93 } $ & $ 1.42 _{ -0.12 } ^{ + 0.14 } $ & $ 0.68 _{ -0.08 } ^{ + 0.09 } $ & --- \\  
    NGC4010 & $ -10.623 $ & $ -0.053 _{ -0.047 } ^{ + 0.050 } $ & $ 0.039 _{ -0.005 } ^{ + 0.007 } $ & $ 16.33 _{ -1.80 } ^{ + 2.03 } $ & $ 88.80 _{ -0.91 } ^{ + 0.75 } $ & $ 1.41 _{ -0.12 } ^{ + 0.14 } $ & $ 0.36 _{ -0.06 } ^{ + 0.07 } $ & --- \\  
    NGC4013 & $ -10.686  $ & $ -0.046 _{ -0.015 } ^{ + 0.014 } $ & $ 0.045 _{ -0.006 } ^{ + 0.002 } $ & $ 14.40 _{ -1.24 } ^{ + 1.34 } $ & $ 88.80 _{ -0.91 } ^{ + 0.75 } $ & $ 1.38 _{ -0.12 } ^{ + 0.13 } $ & $ 0.48 _{ -0.07 } ^{ + 0.08 } $ & $ 0.82 _{ -0.16 } ^{ + 0.20 } $  \\  
    NGC4051 & $ -10.191 $ & $ 0.338 _{ -0.155 } ^{ + 0.113 } $ & $ 0.042 _{ -0.005 } ^{ + 0.008 } $ & $ 17.13 _{ -1.83 } ^{ + 2.04 } $ & $ 49.18 _{ -2.78 } ^{ + 2.82 } $ & $ 1.39 _{ -0.12 } ^{ + 0.13 } $ & $ 0.48 _{ -0.07 } ^{ + 0.08 } $ & --- \\  
    NGC4068 & $ -11.173 $ & $ 0.344 _{ -0.117 } ^{ + 0.103 } $ & $ 0.033 _{ -0.003 } ^{ + 0.006 } $ & $ 4.36 _{ -0.21 } ^{ + 0.22 } $ & $ 47.34 _{ -4.59 } ^{ + 4.47 } $ & $ 1.38 _{ -0.12 } ^{ + 0.13 } $ & $ 0.43 _{ -0.08 } ^{ + 0.09 } $ & --- \\  
    NGC4085 & $ -10.142 $ & $ -0.146 _{ -0.086 } ^{ + 0.106 } $ & $ 0.044 _{ -0.006 } ^{ + 0.003 } $ & $ 15.26 _{ -1.69 } ^{ + 1.90 } $ & $ 81.88 _{ -2.02 } ^{ + 2.01 } $ & $ 1.39 _{ -0.12 } ^{ + 0.13 } $ & $ 0.33 _{ -0.05 } ^{ + 0.06 } $ & --- \\  
    NGC4088 & $ -10.303 $ & $ 0.044 _{ -0.048 } ^{ + 0.053 } $ & $ 0.042 _{ -0.005 } ^{ + 0.008 } $ & $ 14.87 _{ -1.51 } ^{ + 1.68 } $ & $ 68.71 _{ -2.00 } ^{ + 2.00 } $ & $ 1.40 _{ -0.12 } ^{ + 0.13 } $ & $ 0.35 _{ -0.05 } ^{ + 0.05 } $ & --- \\  
    NGC4100 & $ -10.442 $ & $ 0.092 _{ -0.027 } ^{ + 0.028 } $ & $ 0.042 _{ -0.007 } ^{ + 0.005 } $ & $ 19.54 _{ -1.92 } ^{ + 2.14 } $ & $ 73.69 _{ -1.95 } ^{ + 1.95 } $ & $ 1.40 _{ -0.12 } ^{ + 0.13 } $ & $ 0.57 _{ -0.07 } ^{ + 0.08 } $ & --- \\  
    NGC4138 & $ -10.020  $ & $ 0.124 _{ -0.080 } ^{ + 0.104 } $ & $ 0.046 _{ -0.007 } ^{ + 0.004 } $ & $ 18.77 _{ -1.90 } ^{ + 2.12 } $ & $ 54.37 _{ -2.76 } ^{ + 2.79 } $ & $ 1.40 _{ -0.12 } ^{ + 0.14 } $ & $ 0.56 _{ -0.09 } ^{ + 0.11 } $ & $ 0.68 _{ -0.14 } ^{ + 0.17 } $  \\  
    NGC4157 & $ -10.510  $ & $ -0.016 _{ -0.030 } ^{ + 0.031 } $ & $ 0.044 _{ -0.005 } ^{ + 0.006 } $ & $ 15.23 _{ -1.46 } ^{ + 1.61 } $ & $ 81.93 _{ -3.00 } ^{ + 2.98 } $ & $ 1.40 _{ -0.12 } ^{ + 0.13 } $ & $ 0.42 _{ -0.06 } ^{ + 0.06 } $ & $ 0.65 _{ -0.13 } ^{ + 0.16 } $  \\  
    NGC4183 & $ -10.877 $ & $ 0.115 _{ -0.033 } ^{ + 0.039 } $ & $ 0.042 _{ -0.003 } ^{ + 0.009 } $ & $ 18.22 _{ -1.80 } ^{ + 2.04 } $ & $ 82.23 _{ -1.97 } ^{ + 1.97 } $ & $ 1.33 _{ -0.12 } ^{ + 0.13 } $ & $ 0.69 _{ -0.09 } ^{ + 0.11 } $ & --- \\  
    NGC4217 & $ -10.173  $ & $ -0.135 _{ -0.039 } ^{ + 0.039 } $ & $ 0.041 _{ -0.004 } ^{ + 0.005 } $ & $ 15.51 _{ -1.40 } ^{ + 1.53 } $ & $ 85.95 _{ -1.97 } ^{ + 1.90 } $ & $ 1.41 _{ -0.12 } ^{ + 0.14 } $ & $ 0.87 _{ -0.15 } ^{ + 0.18 } $ & $ 0.23 _{ -0.03 } ^{ + 0.03 } $  \\  
    NGC4559 & $ -10.832 $ & $ 0.029 _{ -0.036 } ^{ + 0.040 } $ & $ 0.040 _{ -0.003 } ^{ + 0.009 } $ & $ 7.46 _{ -1.04 } ^{ + 1.26 } $ & $ 67.19 _{ -0.99 } ^{ + 0.99 } $ & $ 1.38 _{ -0.12 } ^{ + 0.13 } $ & $ 0.46 _{ -0.07 } ^{ + 0.09 } $ & --- \\  
    NGC5005 & $ -9.452  $ & $ -0.080 _{ -0.170 } ^{ + 0.252 } $ & $ 0.050 _{ -0.003 } ^{ + 0.007 } $ & $ 16.19 _{ -1.16 } ^{ + 1.24 } $ & $ 68.17 _{ -1.96 } ^{ + 1.97 } $ & $ 1.42 _{ -0.12 } ^{ + 0.14 } $ & $ 0.50 _{ -0.08 } ^{ + 0.08 } $ & $ 0.54 _{ -0.07 } ^{ + 0.08 } $  \\  
    NGC5033 & $ -10.737  $ & $ 0.104 _{ -0.012 } ^{ + 0.013 } $ & $ 0.050 _{ -0.003 } ^{ + 0.008 } $ & $ 23.50 _{ -1.82 } ^{ + 1.96 } $ & $ 66.26 _{ -0.99 } ^{ + 0.99 } $ & $ 1.45 _{ -0.12 } ^{ + 0.13 } $ & $ 0.43 _{ -0.05 } ^{ + 0.06 } $ & $ 0.28 _{ -0.04 } ^{ + 0.04 } $  \\  
    NGC5055 & $ -10.536 $ & $ 0.054 _{ -0.004 } ^{ + 0.005 } $ & $ 0.040 _{ -0.003 } ^{ + 0.006 } $ & $ 9.82 _{ -0.28 } ^{ + 0.29 } $ & $ 59.40 _{ -2.19 } ^{ + 2.37 } $ & $ 1.50 _{ -0.13 } ^{ + 0.14 } $ & $ 0.31 _{ -0.02 } ^{ + 0.02 } $ & --- \\  
    NGC5371 & $ -9.751 $ & $ 0.284 _{ -0.053 } ^{ + 0.054 } $ & $ 0.035 _{ -0.005 } ^{ + 0.009 } $ & $ 16.48 _{ -2.16 } ^{ + 2.50 } $ & $ 52.12 _{ -2.00 } ^{ + 2.00 } $ & $ 1.36 _{ -0.12 } ^{ + 0.13 } $ & $ 1.38 _{ -0.20 } ^{ + 0.23 } $ & --- \\  
    NGC5585 & $ -10.826 $ & $ -0.079 _{ -0.030 } ^{ + 0.030 } $ & $ 0.040 _{ -0.004 } ^{ + 0.005 } $ & $ 5.02 _{ -0.70 } ^{ + 0.81 } $ & $ 51.78 _{ -1.94 } ^{ + 1.95 } $ & $ 1.38 _{ -0.12 } ^{ + 0.13 } $ & $ 0.36 _{ -0.06 } ^{ + 0.07 } $ & --- \\  
    NGC5907 & $ -10.578 $ & $ 0.095 _{ -0.014 } ^{ + 0.015 } $ & $ 0.038 _{ -0.004 } ^{ + 0.006 } $ & $ 16.09 _{ -0.76 } ^{ + 0.80 } $ & $ 87.51 _{ -1.87 } ^{ + 1.54 } $ & $ 1.31 _{ -0.11 } ^{ + 0.12 } $ & $ 0.65 _{ -0.04 } ^{ + 0.04 } $ & --- \\  
    NGC5985 & $ -10.293  $ & $ 0.191 _{ -0.035 } ^{ + 0.039 } $ & $ 0.029 _{ -0.005 } ^{ + 0.001 } $ & $ 72.68 _{ -8.02 } ^{ + 8.97 } $ & $ 62.12 _{ -1.89 } ^{ + 1.90 } $ & $ 1.34 _{ -0.11 } ^{ + 0.13 } $ & $ 0.43 _{ -0.06 } ^{ + 0.07 } $ & $ 1.85 _{ -0.23 } ^{ + 0.26 } $  \\  
    NGC6015 & $ -10.153 $ & $ -0.088 _{ -0.023 } ^{ + 0.022 } $ & $ 0.036 _{ -0.004 } ^{ + 0.003 } $ & $ 8.07 _{ -0.78 } ^{ + 0.86 } $ & $ 60.87 _{ -1.94 } ^{ + 1.95 } $ & $ 1.37 _{ -0.12 } ^{ + 0.13 } $ & $ 1.68 _{ -0.18 } ^{ + 0.21 } $ & --- \\  
    NGC6195 & $ -9.935  $ & $ -0.010 _{ -0.035 } ^{ + 0.036 } $ & $ 0.056 _{ -0.010 } ^{ + 0.008 } $ & $ 110.28 _{ -8.65 } ^{ + 9.45 } $ & $ 59.81 _{ -4.20 } ^{ + 4.44 } $ & $ 1.46 _{ -0.13 } ^{ + 0.14 } $ & $ 0.29 _{ -0.04 } ^{ + 0.05 } $ & $ 0.80 _{ -0.08 } ^{ + 0.09 } $  \\  
    NGC6503 & $ -11.156 $ & $ 0.008 _{ -0.006 } ^{ + 0.006 } $ & $ 0.036 _{ -0.003 } ^{ + 0.004 } $ & $ 6.80 _{ -0.28 } ^{ + 0.29 } $ & $ 75.79 _{ -1.85 } ^{ + 1.85 } $ & $ 1.37 _{ -0.12 } ^{ + 0.13 } $ & $ 0.41 _{ -0.03 } ^{ + 0.03 } $ & --- \\  
    NGC6674 & $ -10.613  $ & $ -0.015 _{ -0.022 } ^{ + 0.019 } $ & $ 0.011 _{ -0.002 } ^{ + 0.003 } $ & $ 37.40 _{ -5.27 } ^{ + 6.20 } $ & $ 52.29 _{ -5.19 } ^{ + 5.50 } $ & $ 1.37 _{ -0.12 } ^{ + 0.13 } $ & $ 1.08 _{ -0.28 } ^{ + 0.33 } $ & $ 1.40 _{ -0.41 } ^{ + 0.59 } $  \\  
    NGC6789 & $ -10.543 $ & $ -0.231 _{ -0.107 } ^{ + 0.106 } $ & $ 0.033 _{ -0.002 } ^{ + 0.005 } $ & $ 3.52 _{ -0.17 } ^{ + 0.18 } $ & $ 46.96 _{ -6.42 } ^{ + 6.61 } $ & $ 1.35 _{ -0.12 } ^{ + 0.13 } $ & $ 0.51 _{ -0.10 } ^{ + 0.13 } $ & --- \\  
    NGC6946 & $ -10.232  $ & $ 0.047 _{ -0.027 } ^{ + 0.026 } $ & $ 0.033 _{ -0.003 } ^{ + 0.004 } $ & $ 4.26 _{ -0.49 } ^{ + 0.55 } $ & $ 41.94 _{ -1.81 } ^{ + 1.82 } $ & $ 1.40 _{ -0.12 } ^{ + 0.13 } $ & $ 0.48 _{ -0.06 } ^{ + 0.07 } $ & $ 0.56 _{ -0.06 } ^{ + 0.07 } $  \\  
    NGC7331 & $ -10.246  $ & $ -0.075 _{ -0.018 } ^{ + 0.017 } $ & $ 0.024 _{ -0.001 } ^{ + 0.003 } $ & $ 12.28 _{ -0.83 } ^{ + 0.89 } $ & $ 74.95 _{ -1.98 } ^{ + 1.98 } $ & $ 1.34 _{ -0.11 } ^{ + 0.12 } $ & $ 0.41 _{ -0.04 } ^{ + 0.04 } $ & $ 0.63 _{ -0.12 } ^{ + 0.14 } $  \\  
    NGC7793 & $ -10.476 $ & $ 0.265 _{ -0.070 } ^{ + 0.087 } $ & $ 0.029 _{ -0.002 } ^{ + 0.008 } $ & $ 3.59 _{ -0.17 } ^{ + 0.18 } $ & $ 69.24 _{ -5.73 } ^{ + 5.98 } $ & $ 1.45 _{ -0.13 } ^{ + 0.14 } $ & $ 0.33 _{ -0.03 } ^{ + 0.04 } $ & --- \\  
    NGC7814 & $ -10.119  $ & $ -0.104 _{ -0.018 } ^{ + 0.018 } $ & $ 0.022 _{ -0.002 } ^{ + 0.002 } $ & $ 14.76 _{ -0.61 } ^{ + 0.64 } $ & $ 89.33 _{ -0.73 } ^{ + 0.47 } $ & $ 1.40 _{ -0.12 } ^{ + 0.14 } $ & $ 0.83 _{ -0.12 } ^{ + 0.13 } $ & $ 0.58 _{ -0.05 } ^{ + 0.05 } $  \\  
   UGC00128 & $ -11.228 $ & $ 0.016 _{ -0.007 } ^{ + 0.007 } $ & $ 0.029 _{ -0.005 } ^{ + 0.002 } $ & $ 49.37 _{ -5.80 } ^{ + 6.76 } $ & $ 52.53 _{ -4.89 } ^{ + 5.65 } $ & $ 1.12 _{ -0.09 } ^{ + 0.10 } $ & $ 1.78 _{ -0.19 } ^{ + 0.21 } $ & --- \\  
   UGC00191 & $ -10.736 $ & $ 0.103 _{ -0.042 } ^{ + 0.056 } $ & $ 0.021 _{ -0.003 } ^{ + 0.002 } $ & $ 16.17 _{ -2.61 } ^{ + 3.28 } $ & $ 48.05 _{ -4.25 } ^{ + 4.39 } $ & $ 1.28 _{ -0.11 } ^{ + 0.12 } $ & $ 0.78 _{ -0.11 } ^{ + 0.13 } $ & --- \\  
   UGC00634 & $ -11.252 $ & $ 0.029 _{ -0.029 } ^{ + 0.034 } $ & $ 0.017 _{ -0.003 } ^{ + 0.006 } $ & $ 29.87 _{ -4.99 } ^{ + 6.19 } $ & $ 41.44 _{ -4.83 } ^{ + 5.38 } $ & $ 1.39 _{ -0.12 } ^{ + 0.13 } $ & $ 0.45 _{ -0.08 } ^{ + 0.10 } $ & --- \\  
   UGC00731 & $ -11.329 $ & $ -0.185 _{ -0.057 } ^{ + 0.390 } $ & $ 0.024 _{ -0.003 } ^{ + 0.002 } $ & $ 5.02 _{ -1.11 } ^{ + 12.66 } $ & $ 56.35 _{ -3.55 } ^{ + 3.65 } $ & $ 1.16 _{ -0.13 } ^{ + 0.14 } $ & $ 0.75 _{ -0.24 } ^{ + 2.32 } $ & --- \\  
   UGC00891 & $ -11.289 $ & $ -0.106 _{ -0.024 } ^{ + 0.021 } $ & $ 0.023 _{ -0.001 } ^{ + 0.003 } $ & $ 5.44 _{ -0.76 } ^{ + 0.90 } $ & $ 56.32 _{ -4.91 } ^{ + 5.03 } $ & $ 1.34 _{ -0.12 } ^{ + 0.13 } $ & $ 0.42 _{ -0.08 } ^{ + 0.10 } $ & --- \\  
   UGC01281 & $ -11.381 $ & $ 0.015 _{ -0.018 } ^{ + 0.020 } $ & $ 0.030 _{ -0.002 } ^{ + 0.004 } $ & $ 5.32 _{ -0.23 } ^{ + 0.24 } $ & $ 89.33 _{ -0.73 } ^{ + 0.47 } $ & $ 1.42 _{ -0.12 } ^{ + 0.13 } $ & $ 0.45 _{ -0.08 } ^{ + 0.09 } $ & --- \\  
   UGC02259 & $ -10.967 $ & $ 0.223 _{ -0.071 } ^{ + 0.099 } $ & $ 0.024 _{ -0.003 } ^{ + 0.002 } $ & $ 16.03 _{ -2.65 } ^{ + 3.25 } $ & $ 44.36 _{ -2.71 } ^{ + 2.74 } $ & $ 1.30 _{ -0.11 } ^{ + 0.12 } $ & $ 0.89 _{ -0.13 } ^{ + 0.16 } $ & --- \\  
 \hline
\end{tabular}
\end{center}
\end{table}

\newpage

\setcounter{table}{1}
\begin{table}
\caption{(continued) Fitted model parameters}
\begin{center}
  \begin{tabular}{lcccccccc} \hline
 galaxy  & $\langle x_0\rangle$   & $e$  & $e_{\rm env}$ & $D$ [Mpc]  & $i$ [$^\circ$] & $\Upsilon_{\rm gas}$ & $\Upsilon_{\rm disk}$  & $\Upsilon_{\rm bulge}$  \\
 \hline
   UGC02487 & $ -10.562  $ & $ 0.100 _{ -0.012 } ^{ + 0.012 } $ & $ 0.043 _{ -0.005 } ^{ + 0.002 } $ & $ 73.77 _{ -8.03 } ^{ + 9.09 } $ & $ 46.14 _{ -3.42 } ^{ + 3.57 } $ & $ 1.49 _{ -0.13 } ^{ + 0.15 } $ & $ 0.58 _{ -0.10 } ^{ + 0.12 } $ & $ 0.59 _{ -0.09 } ^{ + 0.10 } $  \\  
   UGC02885 & $ -10.646  $ & $ 0.009 _{ -0.025 } ^{ + 0.026 } $ & $ 0.050 _{ -0.007 } ^{ + 0.005 } $ & $ 81.67 _{ -6.32 } ^{ + 6.86 } $ & $ 66.70 _{ -3.53 } ^{ + 3.60 } $ & $ 1.44 _{ -0.13 } ^{ + 0.14 } $ & $ 0.44 _{ -0.06 } ^{ + 0.06 } $ & $ 0.92 _{ -0.10 } ^{ + 0.11 } $  \\  
   UGC02916 & $ -9.938  $ & $ 0.295 _{ -0.074 } ^{ + 0.090 } $ & $ 0.041 _{ -0.004 } ^{ + 0.006 } $ & $ 58.36 _{ -5.64 } ^{ + 6.36 } $ & $ 58.52 _{ -3.78 } ^{ + 3.95 } $ & $ 1.39 _{ -0.12 } ^{ + 0.13 } $ & $ 1.10 _{ -0.13 } ^{ + 0.15 } $ & $ 0.43 _{ -0.04 } ^{ + 0.05 } $  \\  
   UGC02953 & $ -9.497  $ & $ -0.006 _{ -0.006 } ^{ + 0.006 } $ & $ 0.026 _{ -0.005 } ^{ + 0.003 } $ & $ 13.51 _{ -0.76 } ^{ + 0.86 } $ & $ 64.56 _{ -3.03 } ^{ + 3.05 } $ & $ 1.51 _{ -0.13 } ^{ + 0.15 } $ & $ 0.57 _{ -0.02 } ^{ + 0.03 } $ & $ 0.58 _{ -0.02 } ^{ + 0.02 } $  \\  
   UGC03205 & $ -9.813  $ & $ 0.004 _{ -0.019 } ^{ + 0.019 } $ & $ 0.012 _{ -0.002 } ^{ + 0.003 } $ & $ 42.44 _{ -4.07 } ^{ + 4.52 } $ & $ 70.75 _{ -3.54 } ^{ + 3.59 } $ & $ 1.33 _{ -0.11 } ^{ + 0.12 } $ & $ 0.63 _{ -0.08 } ^{ + 0.09 } $ & $ 1.30 _{ -0.13 } ^{ + 0.14 } $  \\  
   UGC03546 & $ -10.141  $ & $ 0.020 _{ -0.023 } ^{ + 0.022 } $ & $ 0.015 _{ -0.001 } ^{ + 0.003 } $ & $ 24.38 _{ -2.88 } ^{ + 3.30 } $ & $ 60.82 _{ -4.19 } ^{ + 4.29 } $ & $ 1.40 _{ -0.12 } ^{ + 0.13 } $ & $ 0.58 _{ -0.08 } ^{ + 0.10 } $ & $ 0.43 _{ -0.05 } ^{ + 0.06 } $  \\  
   UGC03580 & $ -10.262  $ & $ -0.045 _{ -0.011 } ^{ + 0.011 } $ & $ 0.023 _{ -0.001 } ^{ + 0.004 } $ & $ 15.18 _{ -1.24 } ^{ + 1.38 } $ & $ 67.11 _{ -3.54 } ^{ + 3.58 } $ & $ 1.51 _{ -0.13 } ^{ + 0.14 } $ & $ 0.47 _{ -0.05 } ^{ + 0.06 } $ & $ 0.15 _{ -0.02 } ^{ + 0.02 } $  \\  
   UGC04278 & $ -11.272 $ & $ -0.164 _{ -0.038 } ^{ + 0.033 } $ & $ 0.032 _{ -0.004 } ^{ + 0.005 } $ & $ 5.89 _{ -0.89 } ^{ + 0.97 } $ & $ 88.01 _{ -2.18 } ^{ + 1.40 } $ & $ 1.38 _{ -0.12 } ^{ + 0.13 } $ & $ 0.37 _{ -0.07 } ^{ + 0.08 } $ & --- \\  
   UGC04325 & $ -10.656 $ & $ 0.354 _{ -0.125 } ^{ + 0.100 } $ & $ 0.031 _{ -0.002 } ^{ + 0.006 } $ & $ 13.64 _{ -2.01 } ^{ + 2.31 } $ & $ 43.99 _{ -2.69 } ^{ + 2.73 } $ & $ 1.25 _{ -0.11 } ^{ + 0.12 } $ & $ 1.07 _{ -0.17 } ^{ + 0.20 } $ & --- \\  
   UGC04483 & $ -11.284 $ & $ 0.153 _{ -0.049 } ^{ + 0.061 } $ & $ 0.032 _{ -0.003 } ^{ + 0.005 } $ & $ 3.37 _{ -0.29 } ^{ + 0.31 } $ & $ 58.97 _{ -2.95 } ^{ + 2.94 } $ & $ 1.37 _{ -0.12 } ^{ + 0.13 } $ & $ 0.48 _{ -0.09 } ^{ + 0.11 } $ & --- \\  
   UGC04499 & $ -11.102 $ & $ 0.121 _{ -0.076 } ^{ + 0.107 } $ & $ 0.035 _{ -0.004 } ^{ + 0.005 } $ & $ 13.43 _{ -2.69 } ^{ + 3.56 } $ & $ 51.41 _{ -2.88 } ^{ + 2.89 } $ & $ 1.38 _{ -0.12 } ^{ + 0.13 } $ & $ 0.49 _{ -0.08 } ^{ + 0.09 } $ & --- \\  
   UGC05005 & $ -11.701 $ & $ 0.149 _{ -0.087 } ^{ + 0.125 } $ & $ 0.022 _{ -0.002 } ^{ + 0.004 } $ & $ 51.72 _{ -8.86 } ^{ + 10.54 } $ & $ 52.66 _{ -8.71 } ^{ + 8.76 } $ & $ 1.42 _{ -0.13 } ^{ + 0.14 } $ & $ 0.39 _{ -0.07 } ^{ + 0.08 } $ & --- \\  
   UGC05253 & $ -9.707  $ & $ 0.058 _{ -0.009 } ^{ + 0.009 } $ & $ 0.027 _{ -0.001 } ^{ + 0.004 } $ & $ 20.87 _{ -1.94 } ^{ + 2.22 } $ & $ 52.15 _{ -2.98 } ^{ + 3.05 } $ & $ 1.40 _{ -0.12 } ^{ + 0.13 } $ & $ 0.30 _{ -0.03 } ^{ + 0.04 } $ & $ 0.39 _{ -0.03 } ^{ + 0.03 } $  \\  
   UGC05414 & $ -11.019 $ & $ 0.083 _{ -0.095 } ^{ + 0.132 } $ & $ 0.032 _{ -0.002 } ^{ + 0.009 } $ & $ 9.03 _{ -2.07 } ^{ + 2.73 } $ & $ 55.52 _{ -2.96 } ^{ + 2.98 } $ & $ 1.40 _{ -0.12 } ^{ + 0.14 } $ & $ 0.42 _{ -0.08 } ^{ + 0.10 } $ & --- \\  
   UGC05716 & $ -11.440 $ & $ 0.091 _{ -0.028 } ^{ + 0.035 } $ & $ 0.040 _{ -0.004 } ^{ + 0.011 } $ & $ 22.21 _{ -3.08 } ^{ + 3.80 } $ & $ 65.58 _{ -7.10 } ^{ + 7.60 } $ & $ 1.25 _{ -0.10 } ^{ + 0.11 } $ & $ 0.86 _{ -0.09 } ^{ + 0.10 } $ & --- \\  
   UGC05721 & $ -10.917 $ & $ 0.048 _{ -0.030 } ^{ + 0.034 } $ & $ 0.033 _{ -0.004 } ^{ + 0.006 } $ & $ 10.10 _{ -1.36 } ^{ + 1.61 } $ & $ 67.74 _{ -4.21 } ^{ + 4.28 } $ & $ 1.39 _{ -0.12 } ^{ + 0.14 } $ & $ 0.51 _{ -0.09 } ^{ + 0.11 } $ & --- \\  
   UGC05750 & $ -11.469 $ & $ 0.151 _{ -0.067 } ^{ + 0.097 } $ & $ 0.024 _{ -0.003 } ^{ + 0.004 } $ & $ 62.77 _{ -10.27 } ^{ + 12.27 } $ & $ 72.30 _{ -7.93 } ^{ + 7.85 } $ & $ 1.37 _{ -0.12 } ^{ + 0.13 } $ & $ 0.55 _{ -0.11 } ^{ + 0.13 } $ & --- \\  
   UGC05764 & $ -11.191 $ & $ 0.390 _{ -0.086 } ^{ + 0.073 } $ & $ 0.033 _{ -0.003 } ^{ + 0.009 } $ & $ 14.73 _{ -1.80 } ^{ + 1.98 } $ & $ 74.46 _{ -6.70 } ^{ + 6.79 } $ & $ 1.10 _{ -0.09 } ^{ + 0.10 } $ & $ 2.66 _{ -0.29 } ^{ + 0.33 } $ & --- \\  
   UGC05829 & $ -11.263 $ & $ 0.128 _{ -0.163 } ^{ + 0.237 } $ & $ 0.035 _{ -0.005 } ^{ + 0.007 } $ & $ 6.81 _{ -1.62 } ^{ + 1.96 } $ & $ 41.90 _{ -9.29 } ^{ + 9.07 } $ & $ 1.27 _{ -0.11 } ^{ + 0.12 } $ & $ 0.69 _{ -0.17 } ^{ + 0.18 } $ & --- \\  
   UGC05918 & $ -11.581 $ & $ 0.046 _{ -0.064 } ^{ + 0.087 } $ & $ 0.036 _{ -0.003 } ^{ + 0.005 } $ & $ 7.47 _{ -1.82 } ^{ + 2.47 } $ & $ 48.80 _{ -4.87 } ^{ + 4.86 } $ & $ 1.33 _{ -0.12 } ^{ + 0.13 } $ & $ 0.59 _{ -0.12 } ^{ + 0.15 } $ & --- \\  
   UGC05986 & $ -10.836 $ & $ -0.016 _{ -0.036 } ^{ + 0.037 } $ & $ 0.033 _{ -0.002 } ^{ + 0.009 } $ & $ 12.87 _{ -1.97 } ^{ + 2.34 } $ & $ 88.04 _{ -2.13 } ^{ + 1.37 } $ & $ 1.49 _{ -0.13 } ^{ + 0.15 } $ & $ 0.37 _{ -0.06 } ^{ + 0.08 } $ & --- \\  
   UGC06399 & $ -11.019 $ & $ -0.002 _{ -0.044 } ^{ + 0.051 } $ & $ 0.036 _{ -0.003 } ^{ + 0.007 } $ & $ 18.58 _{ -2.21 } ^{ + 2.52 } $ & $ 75.24 _{ -1.98 } ^{ + 1.98 } $ & $ 1.37 _{ -0.12 } ^{ + 0.13 } $ & $ 0.54 _{ -0.10 } ^{ + 0.12 } $ & --- \\  
   UGC06446 & $ -11.165 $ & $ 0.154 _{ -0.067 } ^{ + 0.093 } $ & $ 0.035 _{ -0.003 } ^{ + 0.006 } $ & $ 17.45 _{ -3.02 } ^{ + 3.84 } $ & $ 54.18 _{ -2.78 } ^{ + 2.79 } $ & $ 1.28 _{ -0.11 } ^{ + 0.12 } $ & $ 0.91 _{ -0.12 } ^{ + 0.15 } $ & --- \\  
   UGC06614 & $ -10.336  $ & $ -0.066 _{ -0.036 } ^{ + 0.035 } $ & $ 0.020 _{ -0.002 } ^{ + 0.003 } $ & $ 82.51 _{ -7.43 } ^{ + 8.17 } $ & $ 31.00 _{ -2.82 } ^{ + 3.15 } $ & $ 1.42 _{ -0.12 } ^{ + 0.14 } $ & $ 0.47 _{ -0.09 } ^{ + 0.11 } $ & $ 0.57 _{ -0.11 } ^{ + 0.13 } $  \\  
   UGC06667 & $ -11.287 $ & $ -0.117 _{ -0.023 } ^{ + 0.022 } $ & $ 0.037 _{ -0.002 } ^{ + 0.007 } $ & $ 15.64 _{ -1.66 } ^{ + 1.82 } $ & $ 88.80 _{ -0.91 } ^{ + 0.75 } $ & $ 1.31 _{ -0.11 } ^{ + 0.12 } $ & $ 0.52 _{ -0.11 } ^{ + 0.14 } $ & --- \\  
   UGC06786 & $ -10.094  $ & $ -0.028 _{ -0.012 } ^{ + 0.012 } $ & $ 0.053 _{ -0.007 } ^{ + 0.002 } $ & $ 46.17 _{ -3.99 } ^{ + 4.36 } $ & $ 68.02 _{ -2.70 } ^{ + 2.70 } $ & $ 1.49 _{ -0.13 } ^{ + 0.14 } $ & $ 0.36 _{ -0.04 } ^{ + 0.05 } $ & $ 0.42 _{ -0.04 } ^{ + 0.04 } $  \\  
   UGC06787 & $ -10.632  $ & $ 0.302 _{ -0.025 } ^{ + 0.029 } $ & $ 0.035 _{ -0.003 } ^{ + 0.006 } $ & $ 106.11 _{ -7.52 } ^{ + 8.22 } $ & $ 72.66 _{ -2.50 } ^{ + 2.51 } $ & $ 3.25 _{ -0.25 } ^{ + 0.26 } $ & $ 0.17 _{ -0.01 } ^{ + 0.01 } $ & $ 0.08 _{ -0.01 } ^{ + 0.01 } $  \\  
   UGC06818 & $ -11.263 $ & $ -0.001 _{ -0.035 } ^{ + 0.040 } $ & $ 0.041 _{ -0.006 } ^{ + 0.005 } $ & $ 15.71 _{ -1.93 } ^{ + 2.23 } $ & $ 74.65 _{ -3.03 } ^{ + 3.04 } $ & $ 1.42 _{ -0.13 } ^{ + 0.14 } $ & $ 0.31 _{ -0.06 } ^{ + 0.07 } $ & --- \\  
   UGC06917 & $ -10.831 $ & $ 0.001 _{ -0.043 } ^{ + 0.047 } $ & $ 0.038 _{ -0.004 } ^{ + 0.003 } $ & $ 17.98 _{ -1.97 } ^{ + 2.23 } $ & $ 56.50 _{ -1.95 } ^{ + 1.95 } $ & $ 1.36 _{ -0.12 } ^{ + 0.13 } $ & $ 0.55 _{ -0.08 } ^{ + 0.09 } $ & --- \\  
   UGC06923 & $ -10.790 $ & $ 0.042 _{ -0.072 } ^{ + 0.092 } $ & $ 0.052 _{ -0.008 } ^{ + 0.007 } $ & $ 17.32 _{ -2.06 } ^{ + 2.36 } $ & $ 65.05 _{ -2.00 } ^{ + 1.99 } $ & $ 1.38 _{ -0.12 } ^{ + 0.13 } $ & $ 0.45 _{ -0.08 } ^{ + 0.10 } $ & --- \\  
   UGC06930 & $ -11.033 $ & $ 0.259 _{ -0.106 } ^{ + 0.130 } $ & $ 0.034 _{ -0.002 } ^{ + 0.007 } $ & $ 18.30 _{ -2.10 } ^{ + 2.35 } $ & $ 38.76 _{ -3.80 } ^{ + 3.84 } $ & $ 1.36 _{ -0.12 } ^{ + 0.13 } $ & $ 0.58 _{ -0.09 } ^{ + 0.11 } $ & --- \\  
   UGC06983 & $ -10.952 $ & $ 0.059 _{ -0.036 } ^{ + 0.040 } $ & $ 0.051 _{ -0.009 } ^{ + 0.007 } $ & $ 19.85 _{ -2.06 } ^{ + 2.31 } $ & $ 49.43 _{ -0.99 } ^{ + 0.98 } $ & $ 1.33 _{ -0.11 } ^{ + 0.13 } $ & $ 0.77 _{ -0.10 } ^{ + 0.11 } $ & --- \\  
   UGC07089 & $ -11.166 $ & $ 0.102 _{ -0.055 } ^{ + 0.073 } $ & $ 0.042 _{ -0.001 } ^{ + 0.006 } $ & $ 17.11 _{ -2.12 } ^{ + 2.44 } $ & $ 80.16 _{ -2.98 } ^{ + 2.97 } $ & $ 1.40 _{ -0.12 } ^{ + 0.14 } $ & $ 0.40 _{ -0.07 } ^{ + 0.09 } $ & --- \\  
   UGC07125 & $ -11.386 $ & $ 0.132 _{ -0.050 } ^{ + 0.075 } $ & $ 0.052 _{ -0.008 } ^{ + 0.010 } $ & $ 13.56 _{ -2.31 } ^{ + 3.20 } $ & $ 87.98 _{ -2.19 } ^{ + 1.42 } $ & $ 1.27 _{ -0.11 } ^{ + 0.12 } $ & $ 0.69 _{ -0.09 } ^{ + 0.11 } $ & --- \\  
   UGC07151 & $ -10.712 $ & $ 0.163 _{ -0.056 } ^{ + 0.071 } $ & $ 0.037 _{ -0.003 } ^{ + 0.007 } $ & $ 6.97 _{ -0.32 } ^{ + 0.34 } $ & $ 88.04 _{ -2.14 } ^{ + 1.38 } $ & $ 1.35 _{ -0.12 } ^{ + 0.13 } $ & $ 0.71 _{ -0.09 } ^{ + 0.10 } $ & --- \\  
   UGC07232 & $ -10.680 $ & $ -0.024 _{ -0.082 } ^{ + 0.104 } $ & $ 0.033 _{ -0.003 } ^{ + 0.007 } $ & $ 2.82 _{ -0.16 } ^{ + 0.17 } $ & $ 59.49 _{ -4.97 } ^{ + 4.99 } $ & $ 1.37 _{ -0.12 } ^{ + 0.13 } $ & $ 0.46 _{ -0.09 } ^{ + 0.12 } $ & --- \\  
   UGC07261 & $ -11.097 $ & $ 0.258 _{ -0.143 } ^{ + 0.152 } $ & $ 0.044 _{ -0.004 } ^{ + 0.011 } $ & $ 12.39 _{ -2.51 } ^{ + 3.13 } $ & $ 41.87 _{ -6.43 } ^{ + 6.87 } $ & $ 1.36 _{ -0.12 } ^{ + 0.13 } $ & $ 0.49 _{ -0.08 } ^{ + 0.10 } $ & --- \\  
   UGC07323 & $ -10.898 $ & $ 0.150 _{ -0.140 } ^{ + 0.180 } $ & $ 0.037 _{ -0.003 } ^{ + 0.008 } $ & $ 8.41 _{ -1.91 } ^{ + 2.19 } $ & $ 48.55 _{ -2.90 } ^{ + 2.89 } $ & $ 1.41 _{ -0.12 } ^{ + 0.13 } $ & $ 0.43 _{ -0.08 } ^{ + 0.10 } $ & --- \\  
   UGC07399 & $ -10.920 $ & $ -0.024 _{ -0.040 } ^{ + 0.041 } $ & $ 0.040 _{ -0.005 } ^{ + 0.007 } $ & $ 14.30 _{ -2.15 } ^{ + 2.56 } $ & $ 57.55 _{ -2.79 } ^{ + 2.83 } $ & $ 1.38 _{ -0.12 } ^{ + 0.13 } $ & $ 0.61 _{ -0.11 } ^{ + 0.13 } $ & --- \\  
   UGC07524 & $ -11.181 $ & $ 0.169 _{ -0.052 } ^{ + 0.064 } $ & $ 0.037 _{ -0.004 } ^{ + 0.006 } $ & $ 4.73 _{ -0.23 } ^{ + 0.24 } $ & $ 49.80 _{ -2.77 } ^{ + 2.79 } $ & $ 1.24 _{ -0.11 } ^{ + 0.11 } $ & $ 0.90 _{ -0.12 } ^{ + 0.14 } $ & --- \\  
   UGC07559$^{\rm a}$ &  ---  & $ 0.229 _{ -0.066 } ^{ + 0.092 } $ & $ 0.036 _{ -0.004 } ^{ + 0.007 } $ & $ 4.98 _{ -0.24 } ^{ + 0.25 } $ & $ 61.79 _{ -2.96 } ^{ + 2.94 } $ & $ 1.37 _{ -0.12 } ^{ + 0.13 } $ & $ 0.48 _{ -0.10 } ^{ + 0.12 } $ & --- \\  
   UGC07577$^{\rm a}$ &  ---  & $ 0.429 _{ -0.079 } ^{ + 0.051 } $ & $ 0.033 _{ -0.003 } ^{ + 0.007 } $ & $ 2.55 _{ -0.12 } ^{ + 0.13 } $ & $ 62.77 _{ -2.93 } ^{ + 2.95 } $ & $ 1.32 _{ -0.11 } ^{ + 0.12 } $ & $ 0.42 _{ -0.07 } ^{ + 0.08 } $ & --- \\  
   UGC07603 & $ -10.997 $ & $ -0.068 _{ -0.034 } ^{ + 0.033 } $ & $ 0.036 _{ -0.004 } ^{ + 0.008 } $ & $ 4.69 _{ -0.69 } ^{ + 0.81 } $ & $ 78.38 _{ -2.94 } ^{ + 2.94 } $ & $ 1.39 _{ -0.12 } ^{ + 0.13 } $ & $ 0.44 _{ -0.08 } ^{ + 0.10 } $ & --- \\  
   UGC07690 & $ -10.824 $ & $ 0.249 _{ -0.111 } ^{ + 0.136 } $ & $ 0.037 _{ -0.003 } ^{ + 0.008 } $ & $ 8.91 _{ -1.60 } ^{ + 1.92 } $ & $ 45.30 _{ -4.18 } ^{ + 4.32 } $ & $ 1.36 _{ -0.12 } ^{ + 0.13 } $ & $ 0.53 _{ -0.09 } ^{ + 0.12 } $ & --- \\  
   UGC07866$^{\rm a}$ &  ---  & $ 0.230 _{ -0.094 } ^{ + 0.124 } $ & $ 0.036 _{ -0.003 } ^{ + 0.006 } $ & $ 4.58 _{ -0.22 } ^{ + 0.23 } $ & $ 47.67 _{ -4.64 } ^{ + 4.57 } $ & $ 1.35 _{ -0.12 } ^{ + 0.13 } $ & $ 0.53 _{ -0.10 } ^{ + 0.13 } $ & --- \\  
 \hline
\end{tabular}
\end{center}
\end{table}

\newpage

\setcounter{table}{1}
\begin{table}
\caption{(continued) Fitted model parameters}
\begin{center}
  \begin{tabular}{lcccccccc} \hline
 galaxy  & $\langle x_0\rangle$  & $e$  & $e_{\rm env}$ & $D$ [Mpc]  & $i$ [$^\circ$] & $\Upsilon_{\rm gas}$ & $\Upsilon_{\rm disk}$  & $\Upsilon_{\rm bulge}$  \\
 \hline
   UGC08286 & $ -10.832 $ & $ 0.021 _{ -0.014 } ^{ + 0.015 } $ & $ 0.039 _{ -0.003 } ^{ + 0.006 } $ & $ 6.60 _{ -0.20 } ^{ + 0.21 } $ & $ 88.10 _{ -2.07 } ^{ + 1.34 } $ & $ 1.32 _{ -0.11 } ^{ + 0.12 } $ & $ 1.14 _{ -0.08 } ^{ + 0.09 } $ & --- \\  
   UGC08490 & $ -11.162 $ & $ 0.035 _{ -0.015 } ^{ + 0.016 } $ & $ 0.036 _{ -0.004 } ^{ + 0.006 } $ & $ 5.21 _{ -0.43 } ^{ + 0.47 } $ & $ 55.39 _{ -2.53 } ^{ + 2.57 } $ & $ 1.37 _{ -0.12 } ^{ + 0.13 } $ & $ 0.67 _{ -0.09 } ^{ + 0.11 } $ & --- \\  
   UGC08550 & $ -11.251 $ & $ 0.002 _{ -0.026 } ^{ + 0.027 } $ & $ 0.038 _{ -0.002 } ^{ + 0.008 } $ & $ 6.53 _{ -0.86 } ^{ + 1.01 } $ & $ 88.00 _{ -2.17 } ^{ + 1.40 } $ & $ 1.28 _{ -0.11 } ^{ + 0.12 } $ & $ 0.72 _{ -0.11 } ^{ + 0.13 } $ & --- \\  
   UGC08699 & $ -10.120  $ & $ -0.010 _{ -0.024 } ^{ + 0.023 } $ & $ 0.019 _{ -0.011 } ^{ + 0.034 } $ & $ 37.47 _{ -4.05 } ^{ + 4.57 } $ & $ 80.75 _{ -5.90 } ^{ + 5.30 } $ & $ 1.38 _{ -0.12 } ^{ + 0.13 } $ & $ 0.63 _{ -0.10 } ^{ + 0.12 } $ & $ 0.68 _{ -0.07 } ^{ + 0.08 } $  \\  
   UGC08837 & $ -11.296 $ & $ 0.243 _{ -0.060 } ^{ + 0.079 } $ & $ 0.041 _{ -0.004 } ^{ + 0.006 } $ & $ 7.25 _{ -0.34 } ^{ + 0.36 } $ & $ 80.76 _{ -4.66 } ^{ + 4.46 } $ & $ 1.44 _{ -0.12 } ^{ + 0.13 } $ & $ 0.40 _{ -0.07 } ^{ + 0.08 } $ & --- \\  
   UGC09037 & $ -10.734 $ & $ -0.012 _{ -0.036 } ^{ + 0.038 } $ & $ 0.020 _{ -0.003 } ^{ + 0.001 } $ & $ 73.65 _{ -6.22 } ^{ + 6.81 } $ & $ 63.86 _{ -4.73 } ^{ + 4.85 } $ & $ 1.45 _{ -0.13 } ^{ + 0.14 } $ & $ 0.22 _{ -0.03 } ^{ + 0.03 } $ & --- \\  
   UGC09133 & $ -9.810  $ & $ 0.043 _{ -0.007 } ^{ + 0.007 } $ & $ 0.022 _{ -0.004 } ^{ + 0.005 } $ & $ 35.41 _{ -3.03 } ^{ + 3.55 } $ & $ 64.82 _{ -4.40 } ^{ + 4.56 } $ & $ 1.50 _{ -0.13 } ^{ + 0.15 } $ & $ 0.83 _{ -0.09 } ^{ + 0.09 } $ & $ 0.72 _{ -0.04 } ^{ + 0.04 } $  \\  
   UGC09992$^{\rm a}$ &  ---  & $ 0.361 _{ -0.143 } ^{ + 0.099 } $ & $ 0.036 _{ -0.002 } ^{ + 0.007 } $ & $ 9.34 _{ -1.95 } ^{ + 2.49 } $ & $ 34.77 _{ -5.66 } ^{ + 6.19 } $ & $ 1.34 _{ -0.12 } ^{ + 0.13 } $ & $ 0.50 _{ -0.10 } ^{ + 0.12 } $ & --- \\  
   UGC10310 & $ -11.108 $ & $ 0.285 _{ -0.144 } ^{ + 0.139 } $ & $ 0.036 _{ -0.002 } ^{ + 0.006 } $ & $ 16.04 _{ -3.12 } ^{ + 3.73 } $ & $ 40.49 _{ -4.66 } ^{ + 4.80 } $ & $ 1.31 _{ -0.11 } ^{ + 0.12 } $ & $ 0.66 _{ -0.11 } ^{ + 0.14 } $ & --- \\  
   UGC11455 & $ -9.898 $ & $ -0.033 _{ -0.026 } ^{ + 0.025 } $ & $ 0.025 _{ -0.004 } ^{ + 0.001 } $ & $ 72.36 _{ -7.33 } ^{ + 8.14 } $ & $ 89.33 _{ -0.73 } ^{ + 0.47 } $ & $ 1.42 _{ -0.12 } ^{ + 0.14 } $ & $ 0.46 _{ -0.06 } ^{ + 0.07 } $ & --- \\  
   UGC11557 & $ -10.904 $ & $ 0.350 _{ -0.172 } ^{ + 0.108 } $ & $ 0.024 _{ -0.003 } ^{ + 0.002 } $ & $ 17.98 _{ -3.36 } ^{ + 4.18 } $ & $ 32.83 _{ -5.21 } ^{ + 5.63 } $ & $ 1.40 _{ -0.12 } ^{ + 0.13 } $ & $ 0.35 _{ -0.07 } ^{ + 0.09 } $ & --- \\  
   UGC11820 & $ -11.305 $ & $ -0.014 _{ -0.019 } ^{ + 0.019 } $ & $ 0.024 _{ -0.003 } ^{ + 0.001 } $ & $ 12.15 _{ -2.29 } ^{ + 3.01 } $ & $ 44.26 _{ -5.86 } ^{ + 6.72 } $ & $ 1.20 _{ -0.10 } ^{ + 0.11 } $ & $ 0.98 _{ -0.13 } ^{ + 0.15 } $ & --- \\  
   UGC11914 & $ -9.346  $ & $ -0.396 _{ -0.050 } ^{ + 0.052 } $ & $ 0.023 _{ -0.001 } ^{ + 0.003 } $ & $ 8.80 _{ -1.19 } ^{ + 1.43 } $ & $ 48.85 _{ -3.57 } ^{ + 3.68 } $ & $ 1.42 _{ -0.13 } ^{ + 0.14 } $ & $ 0.30 _{ -0.04 } ^{ + 0.05 } $ & $ 0.89 _{ -0.11 } ^{ + 0.12 } $  \\  
   UGC12506 & $ -10.508 $ & $ 0.241 _{ -0.053 } ^{ + 0.064 } $ & $ 0.027 _{ -0.001 } ^{ + 0.003 } $ & $ 117.17 _{ -9.67 } ^{ + 10.52 } $ & $ 86.11 _{ -3.11 } ^{ + 2.48 } $ & $ 1.43 _{ -0.13 } ^{ + 0.14 } $ & $ 1.04 _{ -0.11 } ^{ + 0.13 } $ & --- \\  
   UGC12632 & $ -11.304 $ & $ 0.277 _{ -0.099 } ^{ + 0.122 } $ & $ 0.026 _{ -0.001 } ^{ + 0.003 } $ & $ 13.03 _{ -2.35 } ^{ + 2.59 } $ & $ 49.17 _{ -2.77 } ^{ + 2.78 } $ & $ 1.25 _{ -0.10 } ^{ + 0.11 } $ & $ 1.05 _{ -0.14 } ^{ + 0.16 } $ & --- \\  
   UGC12732 & $ -11.361 $ & $ 0.136 _{ -0.062 } ^{ + 0.093 } $ & $ 0.023 _{ -0.001 } ^{ + 0.003 } $ & $ 13.22 _{ -2.55 } ^{ + 3.32 } $ & $ 48.25 _{ -4.78 } ^{ + 4.97 } $ & $ 1.26 _{ -0.11 } ^{ + 0.12 } $ & $ 0.86 _{ -0.10 } ^{ + 0.12 } $ & --- \\  
    UGCA442 & $ -11.259 $ & $ -0.050 _{ -0.013 } ^{ + 0.012 } $ & $ 0.031 _{ -0.004 } ^{ + 0.002 } $ & $ 4.20 _{ -0.20 } ^{ + 0.21 } $ & $ 51.17 _{ -3.59 } ^{ + 3.93 } $ & $ 1.29 _{ -0.11 } ^{ + 0.12 } $ & $ 0.45 _{ -0.09 } ^{ + 0.11 } $ & --- \\  
    UGCA444$^{\rm a}$ &  ---  & $ 0.063 _{ -0.023 } ^{ + 0.026 } $ & $ 0.032 _{ -0.004 } ^{ + 0.006 } $ & $ 0.95 _{ -0.05 } ^{ + 0.05 } $ & $ 78.77 _{ -3.88 } ^{ + 3.86 } $ & $ 1.25 _{ -0.11 } ^{ + 0.12 } $ & $ 0.57 _{ -0.12 } ^{ + 0.15 } $ & --- \\  
 \hline
\end{tabular}
\end{center}

{\bf Notes.} See Figure~1 and Section~2.2 for the definition of $x_0$. Here $\langle x_0\rangle$ represents the median of $x_0$ for the rotation velocities with signal-to-noise ratios $> 10$.

$^{\rm a}$ For these galaxies no circular velocities have signal-to-noise ratios $> 10$. These galaxies are not included in our statistical analyses of EFE.

\end{table}

\end{document}